\documentclass[preprint2]{aastex}
\usepackage{color}
\usepackage{txfonts}
\usepackage{breqn}
\usepackage{psfrag,graphicx,natbib}
\shorttitle{The formation of the first galaxies}

\shortauthors{Latif et al.}

\def\Rev. Mod. Phys{Rev.~Mod.~Phys}

\begin{document}

\title {Impact of dust cooling on direct collapse black hole formation}

\author{M. A. Latif \altaffilmark{1,2}, K. Omukai\altaffilmark{3}, M. Habouzit\altaffilmark{1,2}, D.~R.~G.~Schleicher\altaffilmark{4}, M.  Volonteri\altaffilmark{1,2}}
\affil{Sorbonne Universités, UPMC Univ Paris 06, UMR 7095, Institut d'Astrophysique de Paris, F-75014, Paris, France}
\affil{CNRS, UMR 7095, Institut d'Astrophysique de Paris, F-75014, Paris, France}
\affil{Astronomical Institute, Tohoku University, Aoba, Sendai 980-8578, Japan}
\affil{Departamento de Astronomía, Facultad Ciencias Físicas y Matemáticas, Universidad de Concepción, \\
       Av. Esteban Iturra s/n Barrio Universitario, Casilla 160-C, Chile}
\email{latif@iap.fr}
\newcommand{\ch}[1]{\textcolor{black}{\textbf{#1}}}
\date{}

\bibliographystyle{apj}

\begin{abstract}
{
Observations of  quasars at $ z > 6$ suggest the presence of  black holes with a  few  times $\rm 10^9 ~M_{\odot}$. Numerous models have been proposed to explain their existence including the direct collapse which provides massive seeds of  $\rm 10^5~M_{\odot}$.  The isothermal direct collapse requires a strong Lyman-Werner flux to quench $\rm H_2$ formation in massive primordial halos. In this study, we explore the impact of trace amounts of metals and dust  enrichment. We perform three dimensional cosmological simulations for  two  halos of   $\rm > 10^7~M_{\odot}$  with  $\rm Z/Z_{\odot}= 10^{-4}-10^{-6}$ illuminated by an intense Lyman Werner flux of $\rm  J_{21}=10^5$. Our results show that initially the collapse proceeds isothermally  with $\rm T \sim 8000$ K but dust cooling becomes effective at densities of $\rm 10^{8}-10^{12} ~cm^{-3}$ and brings the gas temperature down to a few 100-1000 K for $\rm Z/Z_{\odot} \geq 10^{-6}$.  No gravitationally bound clumps are found in $\rm Z/Z_{\odot} \leq 10^{-5}$ cases  by the end of our simulations in contrast to the case with $\rm Z/Z_{\odot} = 10^{-4}$.  Large inflow rates of $\rm \geq 0.1~M_{\odot}/yr$ are observed for $\rm Z/Z_{\odot} \leq 10^{-5}$ similar to  a zero-metallicity case while for $\rm Z/Z_{\odot} = 10^{-4}$ the inflow rate starts  to decline earlier due to the dust cooling and fragmentation.  For given large inflow rates a central star of  $\rm \sim 10^4~M_{\odot}$ may form  for $\rm Z/Z_{\odot} \leq 10^{-5}$.} 

\end{abstract}



\section{Introduction} \label{sec:intro}
The discovery of high redshift quasars at $z > 6$ reveals the presence of supermassive black holes (SMBHs) of about a few billion solar masses  \citep{Fan2006, Willot2010, MOrtlock2011,Venemans2013,Wu2015,Venemans2015}. The formation of SMBHs  a few hundred million years after the Big Bang presents a challenge for our understanding of early structure formation. The pathways for their assembly include the stellar mass black holes forming from the collapse of Pop III stars \citep{Abel2002,Yoshida08,LatifPopIII13,Hirano2014},  black hole seeds  from the collapse of a dense nuclear cluster either due to  the relativistic instability \citep{Baumgarte1999}, via stellar dynamical process \citep{Omukai2008,Devecchi2009,Katz2015,Yajima2015},  even the core collapse of a cluster consisting of stellar mass black holes \citep{Davies2011ApJ,Lupi2014} or a direct collapse of a protogalactic gas cloud \citep{Loeb1994,Bromm03,Begelman2006,Spaans2006,Latif2013c,Latif2013d,Ferrara14,Inayoshi2014,Choi2014,Shlosman2016}. Details of these models are discussed in dedicated reviews \citep{Volonteri2010,Volonteri2012,Haiman2012}.

The masses of seed black holes from the above mentioned scenarios range from a few hundred to thousand solar masses except the direct collapse model which provides massive seeds of about $\rm 10^5-10^6 ~M_{\odot}$. The  lighter seeds require continuous accretion close to the Eddington limit in order to reach a billion solar masses in a few hundred million years. Therefore, massive seeds forming via the direct collapse are favoured for the assembly of high redshift quasars.  The formation of so-called direct collapse black holes (DCBHs) requires  large inflow rates of about $\rm \geq 0.1 ~M_{\odot}/yr$ \citep{Begelman2010,Hosokawa2013,Schleicher13,Sakurai2015}.  These inflow rates can be obtained either via dynamical processes such as  the 'bars-in-bars' instability (see \cite{Begelman2006} and \cite{Begelman2009}) or thermodynamically due to a large speed of sound ($\rm \dot{M} \sim {c_s^3}/{G} \sim 0.1~M_{\odot}/yr \left( {T}/{8000~K}\right)^{3/2}$,  where $c_s$ is the thermal sound speed). Large inflow rates can be more easily achieved in an isothermal direct collapse which necessitates the presence of a strong Lyman Werner (LW) flux to quench the formation of molecular hydrogen  in metal free halos \citep{Omukai2001, Schleicher10,Shang2010,Latif2014UV}.  The critical value of  the LW flux (J$_{\rm crit}$) above which isothermal collapse occurs depends on the spectra of the stars \citep{Omukai2001, Shang2010,Latif2014UV,Sugimura14,Agarwal2015,Regan2014B,Latif2015a}.  \cite{Latif2015a} computed the strength of J$_{crit}$ for realistic Pop II spectra from three dimensional cosmological simulations and found that it corresponds to a value of J$_{\rm 21}$ of a few times $\rm 10^4$ where $\rm J_{21}=1$ implies a flux of $\rm 10^{-21}~ erg/cm^2/s/Hz/sr$ below the Lyman limit. Another constraint for the isothermal direct collapse  is that  the halos should be metal free as trace amounts of metals and dust can lead to fragmentation and star formation \citep{Omukai2008}.

Numerical simulations show that large inflow rates of about $\rm 0.1-1~M_{\odot}/yr$ are available in the massive primordial halos of $\rm 10^7-10^8~M_{\odot}$ illuminated by a strong LW flux at z=15 and massive objects of $\rm 10^5 ~M_{\odot}$ can be formed within about 100,000 years \citep{Wise2008,Latif2013c,Latif2013d,Regan2014a,Bcerra2014,Inayoshi2014,VanBorm2014}. However,  the value of J$_{crit}$ ($\rm > 10^4 ~ in ~ terms~of~ J_{21}$) required for the formation of isothermal DCBHs is much higher than  the background UV flux and such a flux can only be achieved in the close vicinities of star forming  galaxies \citep{Dijkstra2008,Agarwal2012,Dijksta2014,Visbal2014,Latif2015a,Habouzit2015}. Supernova winds from nearby galaxies can also pollute halos with metals and make the sites for an isothermal direct collapse even more rare \citep{Johnson2013}.  In the context of second generation star formation (Pop II), it has been found that  above a critical value of  metallically ($\rm Z_{crit} = 3 \times 10^{-4} Z_{\odot}$) fragmentation becomes inevitable and leads to the formation of Pop II stars \citep{Omukai2005,GloverJappsen2007,Wise2012,Safranek-Shrader2014,Bovino14,Ritter2014}. The critical value of the metallicity is further reduced  by about two orders of magnitude in the  presence of trace amounts of dust \citep{Schneider2003,Omukai2005,Omukai2008} and the $\rm H_2$ formation on dust grains enhances the cooling \citep{Omukai2001,Cazaux2009,Latif2012}. Numerical simulations show that low mass star formation seems plausible due to the dust cooling occurring at high densities in minihalos with metallicities as low as a few times $\rm 10^{-5}$ solar \citep{Tsuribe2006,Tsuribe2008,Dopcke2013,Smith2015}.

\cite{Omukai2008} explored the thermodynamical properties of gas irradiated by a strong LW flux above $J_{crit}$ and enriched by trace amounts of metals and dust. They found from a one-zone model  that in such conditions dust cooling becomes effective for $\rm Z/Z_{\odot} \geq 5 \times 10^{-6}$ and proposed that a dense stellar cluster can form under these conditions which may later collapse into a black hole of 100-1000 $\rm M_{\odot}$.

In this study, we explore the impact of dust and metal line cooling in massive primordial halos illuminated by a strong background LW flux and polluted by trace amounts of metals and dust. To accomplish this goal, we perform high resolution three dimensional cosmological simulations for two halos of a few times $\rm 10^7 ~M_{\odot}$ by turning on  a LW flux of strength $10^5$ in units of $\rm J_{21}$ at z=30 and presume that they are pre-enriched with  $\rm Z/Z_{\odot}=10^{-6}-10^{-4}$.  Our results suggest that dust cooling occurs even for $\rm Z/Z_{\odot}=10^{-6}$ at densities above $\rm 10^{10} ~cm^{-3}$ but still the conditions are suitable for the formation of massive objects.  This work has important implications for assessing the feasibility of  black hole formation models.

This article is organised as follows. In section 2, we describe numerical methods and  chemical model employed in this work. We present our results in section 3 and discuss its implications for black hole formation in section 4.  Summary of our main results and conclusions are discussed in section 5.

\section{Numerical Methods}
We have performed simulations using the open source code Enzo version 2.4 \citep{Enzocode2014}. Enzo is an adaptive mesh refinement, parallel, grid based cosmological simulation code which has been extensively used to perform high resolution simulations. It uses the message passing interface (MPI) to  achieve scalability and portability on various platforms.  The piece-wise parabolic method (PPM)  is employed to solve  hydrodynamics, the particle-mesh technique (PM) is used to solve dark matter dynamics  and the multi-grid Poisson solver for gravity.

\subsection{Simulation setup}
Our simulations are started with cosmological initial conditions at z=100  generated with the "inits" package available with Enzo. The computational domain  has a size of 1 Mpc/h in comoving units and periodic boundary conditions are adopted both for gravity and hydrodynamics. We use parameters from the WMAP 7 years data to generate initial conditions \citep{Jarosik2011}. Our computational box is centered on the most massive halo forming  at z=15 selected from uniform grid DM only simulations of resolution $\rm 128^3$ particles. We rerun the simulations with a top grid resolution of $\rm 128^3$  grid cells (same number of DM particles) and add two nested refinement levels each with a resolution of  $\rm 128^3$  grid cells and $\rm 128^3$  DM particles. In all, we employ 5767168 particles to solve the DM dynamics which provide us an effective DM resolution of about 600 $\rm M_{\odot}$. We smooth DM particles at about a 2 pc scale to avoid spurious numerical artefacts, see \cite{LatifVolonteri15} for a detailed discussion. We further add 28 dynamical refinement levels during the course of  the simulations in the central 62 kpc comoving region of the box. This approach enables  us to resolve the gravitational collapse down to scales of about 0.1 AU. We employ a fixed Jeans resolution of 32 cells throughout the simulations to resolve turbulent eddies and to fulfil the Truelove criterion which requires a resolution of at-least four cells per Jeans length \citep{TrueLuv,Federrath2011,Latif2013}. Our dynamical refinement criterion further includes the baryonic over density and particle mass resolution. Grid cells are marked for refinement if gas density exceeds  four times the cosmic mean.  Similarly, cells are flagged for refinement if  DM density is above 0.0625 times  $\rho_{DM}r^{\ell \alpha}$  where $\rho_{DM}$ is the dark matter density, r = 2 is the refinement factor, $\ell$ is the refinement level, and $\alpha = -0.3$ makes the refinement super-Lagrangian.  

We select two halos of  $\rm 5.6 \times 10^7 ~M_{\odot}$ and $\rm 3.25 \times 10^7 ~M_{\odot}$. The properties of these halos such as their spins, collapse redshifts and $\rm J_{crit}$ are listed in table 1 of \cite{LatifVolonteri15} as the halo A and the halo C, respectively.  We presume that the halos are enriched with trace amount of metals and dust  by supernova winds from a nearby star forming galaxy. The latter also provides a strong LW flux above the critical threshold required for  the quenching of $\rm H_2$ formation. The prime aim of this study is to explore the impact of dust cooling in the presence of an intense LW flux. We therefore select $\rm J_{21}=10^5$ for which halo collapses isothermally in the absence of dust cooling. At densities above $\rm 10^{-16}~g/cm^{3}$ where dust cooling becomes important the gas cloud  is already self-shielded against external radiation and therefore the value of $\rm J_{crit}$ does not depend on the metal content of the halo. This scenario is particularly relevant for the halos which get polluted while being irradiated  by a strong LW flux, and we specifically explore here the cases with  $\rm Z/Z_{\odot}= 10^{-6},~10^{-5}~and ~10^{-4}$.

 Due to the computational constraints, we stop our simulations when a peak density of $\rm 3 \times 10^{-11} ~g/cm^{3}$ is reached.  \cite{Omukai2008} have shown that dust cooling becomes optically thick at  densities above $\rm \sim 10^{-10}~g/cm^3$ for the metallicities explored here, and thereafter collapse is expected to proceed adiabatically. The transition to the adiabatic regime  during the gravitational collapse of primordial gas has been explored in hydrodynamical simulations \citep{Inayoshi2014, VanBorm2014, Bcerra2014,Latif2016MNRAS}. These studies found that the adiabatic phase stabilises the gravitational collapse on small scales, breaks its self-similarity and a protostar begins to form at this stage.  For dust cooling, \cite{Omukai2008} have shown that the transition to the adiabatic phase occurs at earlier stages of the collapse (at lower densities compared to the primordial gas).  They also considered the sputtering of dust grains in their model finding that the dust temperature remains below the vaporisation temperature in optically thin regime and  once adiabatic phase begins, it exceeds the evaporation temperature but gas cloud is already  optically thick.  Therefore,  thermal evolution remains adiabatic similar to the metal free gas. Further details of our model are described in \cite{Omukai2008}.

In this study, we explore three cases with metallicities  of $\rm Z/Z_{\odot}= 10^{-6}, 10^{-5}~and~10^{-4}$ for each halo and turn on a strong LW flux of strength $\rm J_{21}=10^5$, well above the $\rm J_{crit}$ found in \cite{Latif2015a}. Further details of our chemical network as well as heating and cooling processes are described in the section below.

\subsection{Chemical model}
We employ  the KROME package \citep{Grassi2014} to  solve the chemical and thermal evolution of the gas in cosmological simulations.  The rate equations of $\rm H,~H^+,~ H^-, ~He,~ He^+, ~He^{++},~ H_2, ~H_2^+, ~e^-$ are solved  to study their evolution during the gravitational collapse. We employ a uniform background LW flux of strength $\rm J_{21}=10^5 $ emitted from  a Pop II galaxy which can be mimicked by a radiation  temperature of $\rm 2 \times 10^4 ~K$ (see \cite{Sugimura14,Latif2015a}). Our results do not depend on the choice of the radiation spectrum as long as the flux is well above the critical threshold to dissociate molecular hydrogen (derived by  \cite{Omukai2008,Sugimura14} for the one-zone model and by \cite{Latif2015a} in 3D simulations). The list of chemical reactions is given in table 1 in the appendix of \cite{Latif2015a}.  Our model includes  $\rm H_2$ formation, $\rm H_2$ and $\rm H_2^+$ photo-dissociation, $\rm H^-$ photo-detachment, collisional induced emission and $\rm H_2$ collisional dissociation. In addition to this, we add a reaction for  $\rm  H_2$ formation on dust grain surfaces taken from \cite{Omukai2000}. Various cooling and heating mechanisms such as  cooling due to the collisional excitation, collisional ionisation, radiative recombination, Bremsstrahlung radiation and  chemical cooling/heating from three-body reactions are included in the chemical model. We also employ the $\rm H_2$ self-shielding fitting formula given in \cite{WolocottGreen2011}.

In addition to the cooling and heating processes mentioned above, we include cooling due to  the fine structure lines of CII and OI, cooling and heating by dust grains which is  relevant for $\rm Z/Z_{\odot}= 10^{-6}, 10^{-5}~and ~10^{-4}$.  Our treatment for metal and dust cooling/heating processes comes from the model of  \cite{Omukai2008}, here we  summarise their main features. We do not explicitly include the chemical reactions involving  metals but instead assume that most of the oxygen and carbon are in OI and CII form due to the presence of a strong LW flux.  Carbon has a lower ionisation energy (11.26 eV) than hydrogen and therefore is in CII form while OI remains in the atomic state due to the charge exchange with hydrogen atoms in a neutral medium. The dust grain composition and size distribution is assumed to be similar to the solar neighbourhood  and the amount is scaled with the metallicity of the gas cloud. We assume a dust to gas mass ratio of $\rm 0.01~Z/Z_{\odot}$. Dust grains in the supernova ejecta are more efficient in cooling and $\rm H_2$ formation due to their smaller size and larger surface area \citep{Schneider2003,Omukai2008}. The dust is assumed to be in thermal equilibrium and its temperature ($\rm T_{dust}$) is evaluated separately from the gas temperature at each density and temperature.  The following energy balance equation is solved to compute the $\rm T_{dust}$ \citep{Omukai2008}
\begin{eqnarray}
4 \pi \int \kappa _{a,\nu} B_{\nu} \left( T_{dust} \right) d\nu = \Lambda_{gas \rightarrow dust}  \\ \nonumber
 +  4 \pi \int \kappa _{a,\nu} B_{\nu} \left( T_{CMB} \right) d\nu.
\label{eq0}
\end{eqnarray}
Here $\Lambda_{gas \rightarrow dust}$ is the net gas cooling rate per unit mass due to dust grain collisions, $\kappa _{a,\nu}$ is the dust absorption opacity and $B_{\nu}$ is the black body spectrum.  $\Lambda_{gas \rightarrow dust}$ is given as \citep{Hollenbach79}
\begin{eqnarray}
\Lambda_{gas \rightarrow dust}=5.8 \times 10^{-8} n  \left({T \over 1000}\right)^{1/2} \times  \\ \nonumber
\left[1 - 0.8~exp \left(\frac{-75 K}{T} \right) \right]  (T - T_{dust})  (Z/Z_{\odot})
\label{eq0}
\end{eqnarray}
where n is the gas number density and $\rm T_{dust}$ is the dust temperature. We have set temperature floor at $\rm T_{CMB}$. Further details about the dust properties can be found in section 2.2 of  \cite{Omukai2008}.

\section {Results}
We here present our main results obtained both from a one-zone model  and three dimensional cosmological simulations. 

\subsection{One-zone model}
\begin{figure*}
\includegraphics[scale=1.4]{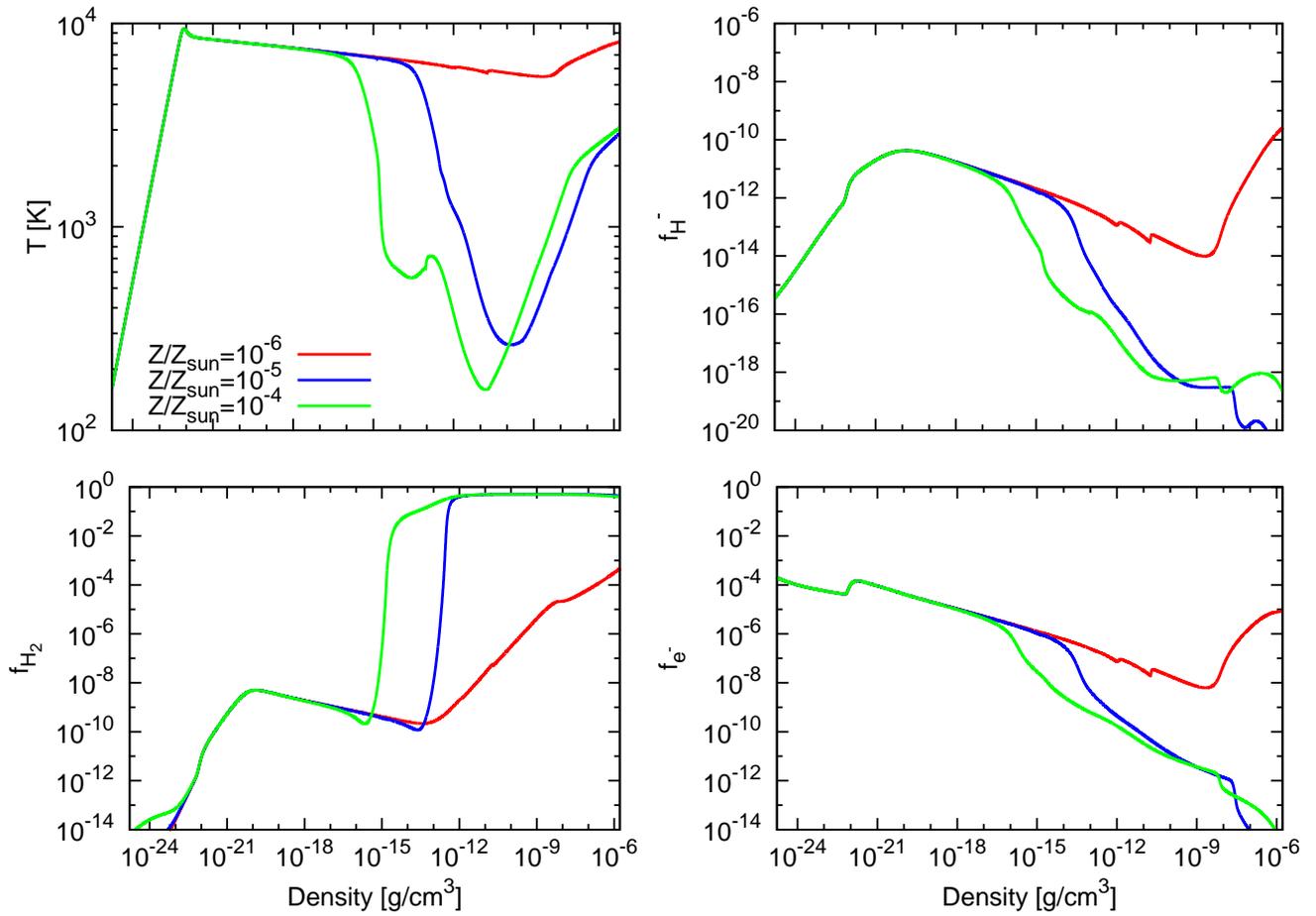}
\caption{Figure shows the temperature, $\rm H_2$, $\rm H^-$ and $\rm e^-$ fractions plotted against the density from one-zone test.  The green, blue and red lines represent $\rm Z/Z_{\odot} =10^{-6}, 10^{-5} ~and~10^{-4}$, respectively.}
\label{fig}
\end{figure*}

To test  the chemical model presented in the previous section, we performed a one-zone test  with a LW flux of strength $\rm J_{21} = 10^5$ (well above the $\rm J_{crit}$ from one-zone model with $\rm T_{rad} = 2 \times 10^4$ K) and  varied the  metallicity  from $\rm Z/Z_{\odot} = 10^{-6} -10^{-4}$. We took an initial temperature of 160 K, a gas density of $\rm \sim 10^{-23} ~g/cm^{3}$ and species abundances of $\rm 2 \times 10^{-6}$  for $\rm e^-$ and $\rm 2 \times 10^{-4}$  for $\rm H_2$.  In Figure \ref{fig}, we show the temperature and the abundances of $\rm H_2$, $\rm H^-$ and $\rm e^-$ as a function of density.  In the presence of a strong LW flux, $\rm H_2$ gets photo-dissociated, its abundance remains below $\rm 10^{-8}$ and is unable to cool the gas. The temperature increases up to about $\rm 10^4 ~K$ where atomic line cooling becomes effective and cools the gas down to 8000 K up to the densities of $\rm 10^{-17} ~g/cm^{3}$. At higher densities for $\rm Z/Z_{\odot} \geq 10^{-5}$, dust cooling comes into play and sharply cools the gas down to a few hundred K. For the case of $\rm Z/Z_{\odot}=10^{-4}$, $\rm H_2$ formation on dust grains becomes effective and its abundances get boosted very quickly.  The small increase in the temperature  at a density of $\rm 10^{-13}~g/cm^{3}$ is due to the chemical heating produced by the formation of $\rm H_2$ on grain surfaces.  Dust cooling continues at high densities up to $\rm 10^{-10}~g/cm^{3}$ and after that the cloud becomes optically thick to dust self-absorption. Beyond this point the thermal evolution proceeds adiabatically and the inclusion of dust cooling breaks the a self-similarly of an isothermal  collapse.  A protostar is expected to form at this stage with a central adiabatic core and is expected to grow rapidly by accretion from the protostellar envelope.

For $\rm Z/Z_{\odot}=10^{-5}$, the dust  cooling becomes effective at densities of $\rm 10^{-14} ~g/cm^{3}$, about two orders of magnitude higher than for $\rm Z/Z_{\odot}=10^{-4}$ case. The temperature sharply declines down to a few hundred K and the $\rm H_2$ fraction gets boosted. The cloud becomes optically thick to dust cooling similar  to the $\rm Z/Z_{\odot}=10^{-4}$  case at  $\rm 10^{-10}~g/cm^{3}$  and the thermal evolution follows the adiabatic equation of state. At  an even lower metallicity of $\rm Z/Z_{\odot}=10^{-6}$, the dust cooling becomes negligible and the thermal evolution follows the atomic cooling track. The $\rm H_2$ fraction is increased up to about a few times $\rm 10^{-4}$. The abundances of $\rm e^-$ and $\rm H^-$ remain very low and are similar at high densities for $\rm Z/Z_{\odot} = 10^{-5}$ and $\rm Z/Z_{\odot}=10^{-4}$ cases. The degree of ionisation is higher for $\rm Z/Z_{\odot}=10^{-6}$. The impact of metal line cooling such as CII and OI is negligible for such metallicities. Overall, our results are in good agreement with \cite{Omukai2008}.

\subsection{3D simulations}
In total, we have performed  six cosmological simulations for two halos  of  $\rm 5.6 \times 10^7 ~M_{\odot}$ (H1) and $\rm 3.25 \times 10^7 ~M_{\odot}$ (H2) for metallicities of  $\rm Z/Z_{\odot}=10^{-4}, ~10^{-5} ~and ~ 10^{-6}$.  The collapse redshifts of the halos are 10.4 and 10.8, respectively, and independent of the dust content. We also performed one additional simulation with zero metallicity for the sake of comparison for H1. The gravitational collapse of the halos was followed from 1 Mpc down to sub-AU scales with the adaptive mesh refinement approach. We assumed that the halos are  irradiated by a strong LW flux of strength $\rm 10^5~in ~terms ~of ~J_{21}$, well above the $\rm J_{crit}$ found  in \cite{Latif2015a}, and the LW flux as well as the fixed metallicities of the above mentioned amount are turned on at z=30.

 \subsubsection{Thermodynamical and dynamical properties}
\begin{figure*}
%
\includegraphics[scale=0.35]{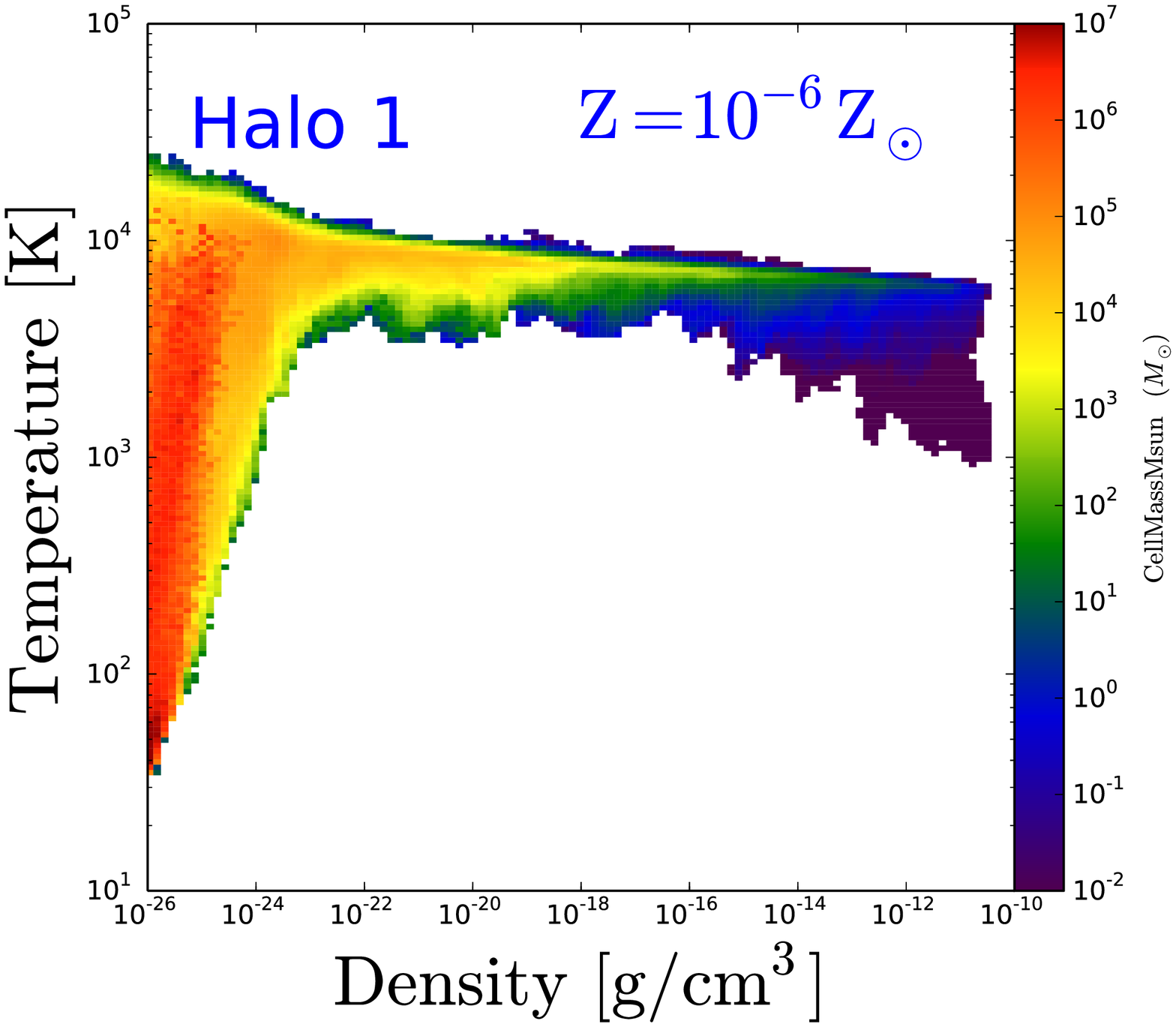} 
\includegraphics[scale=0.35]{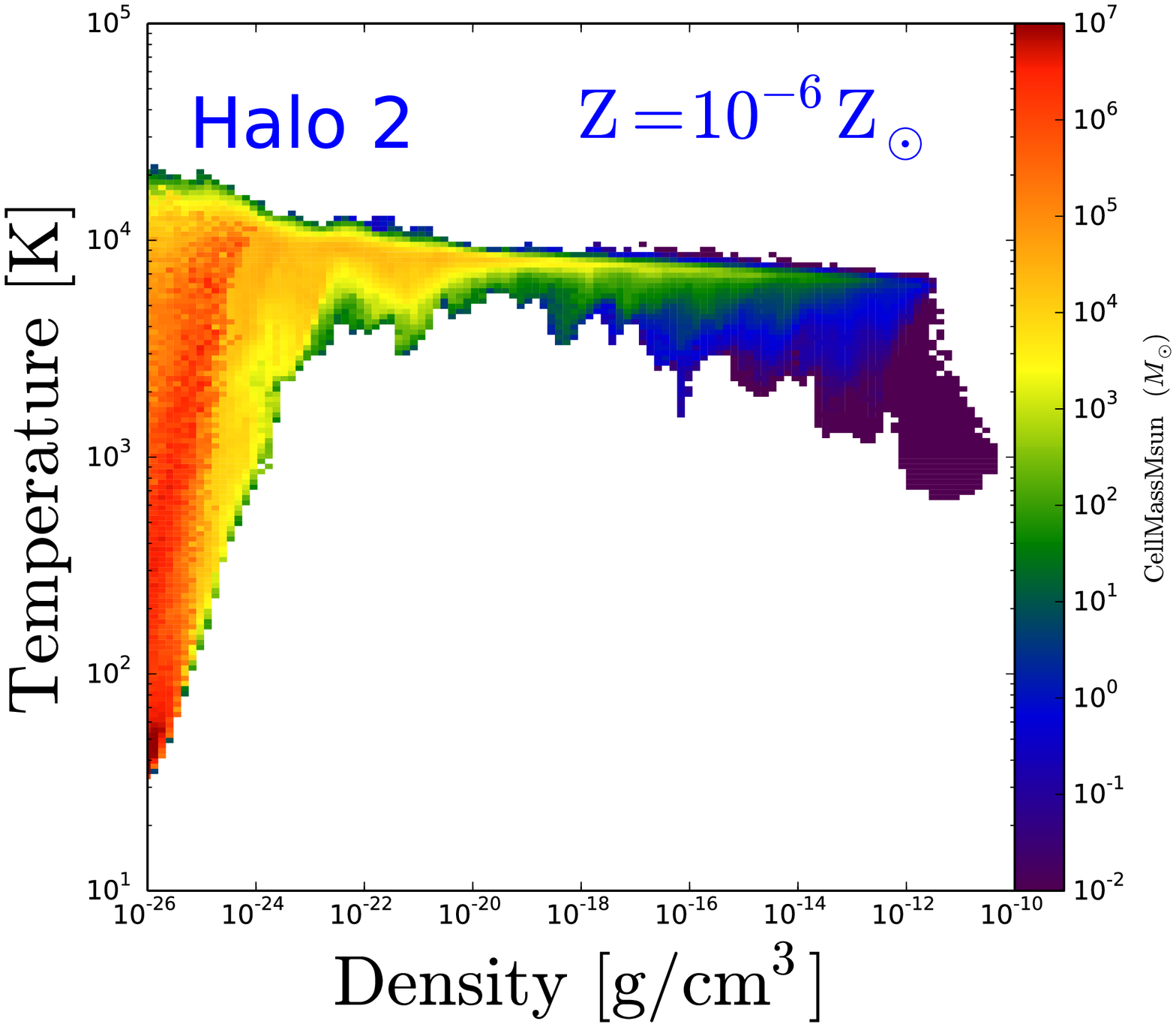} 
\includegraphics[scale=0.35]{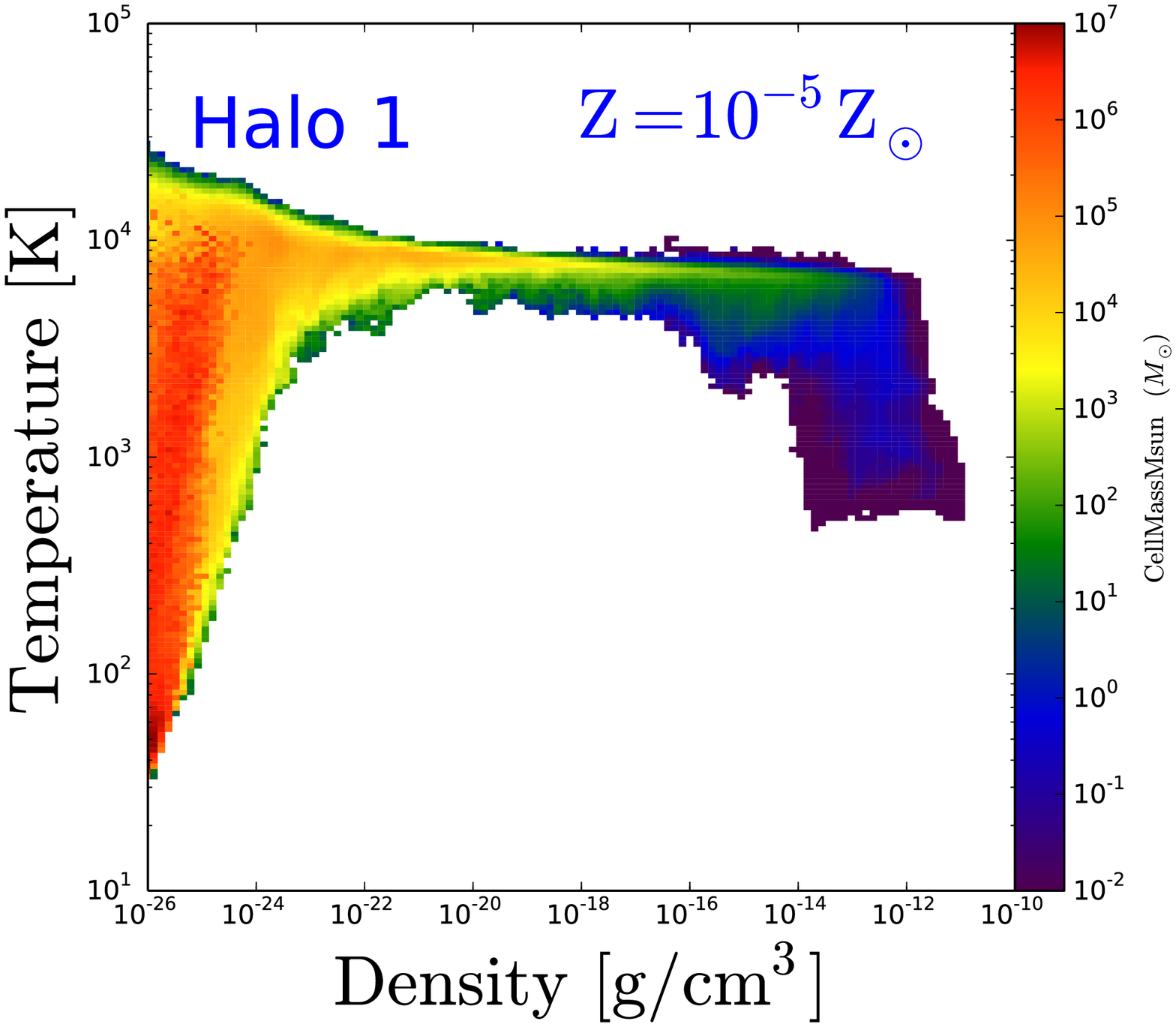} 
\includegraphics[scale=0.35]{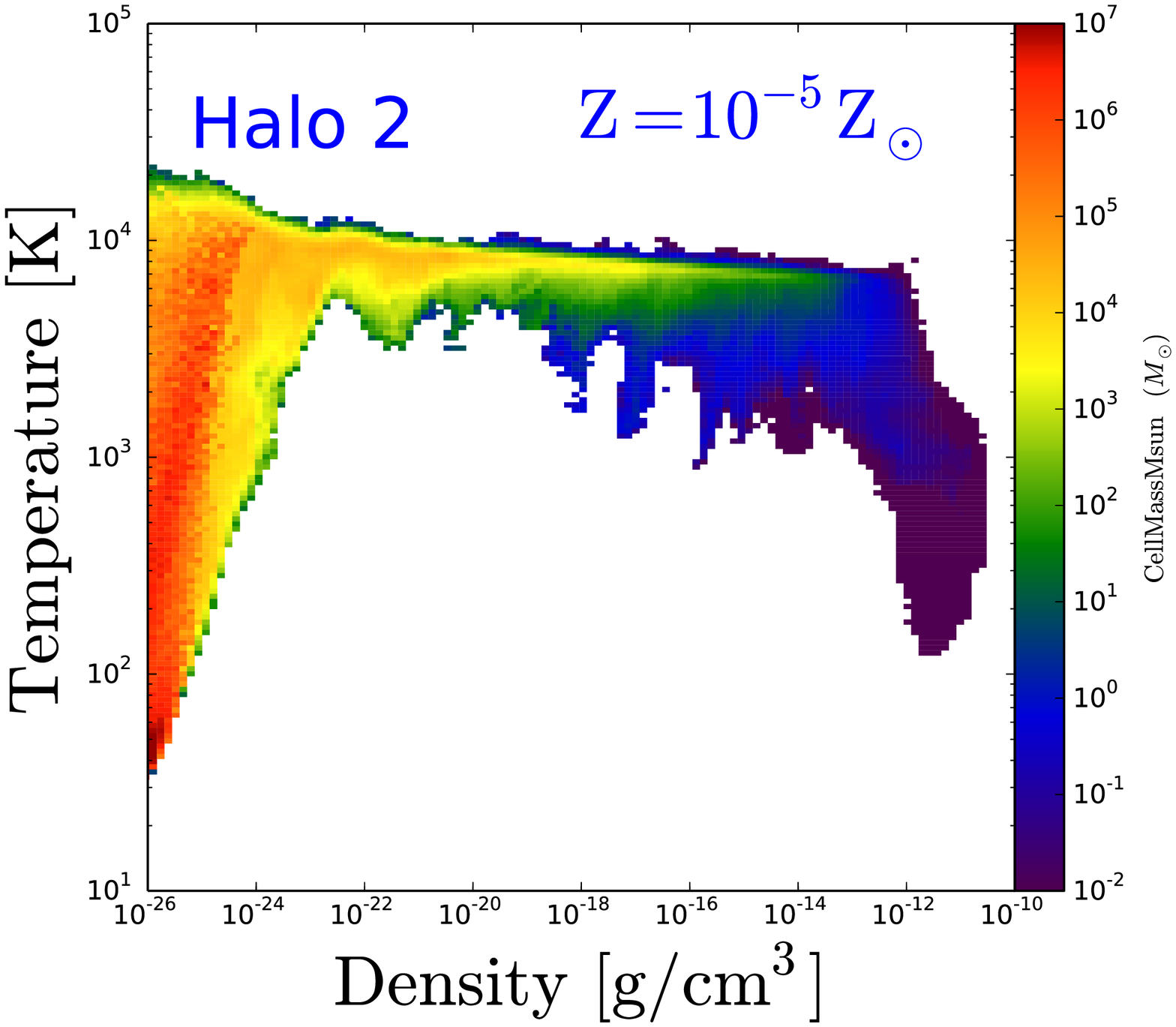} 
\includegraphics[scale=0.35]{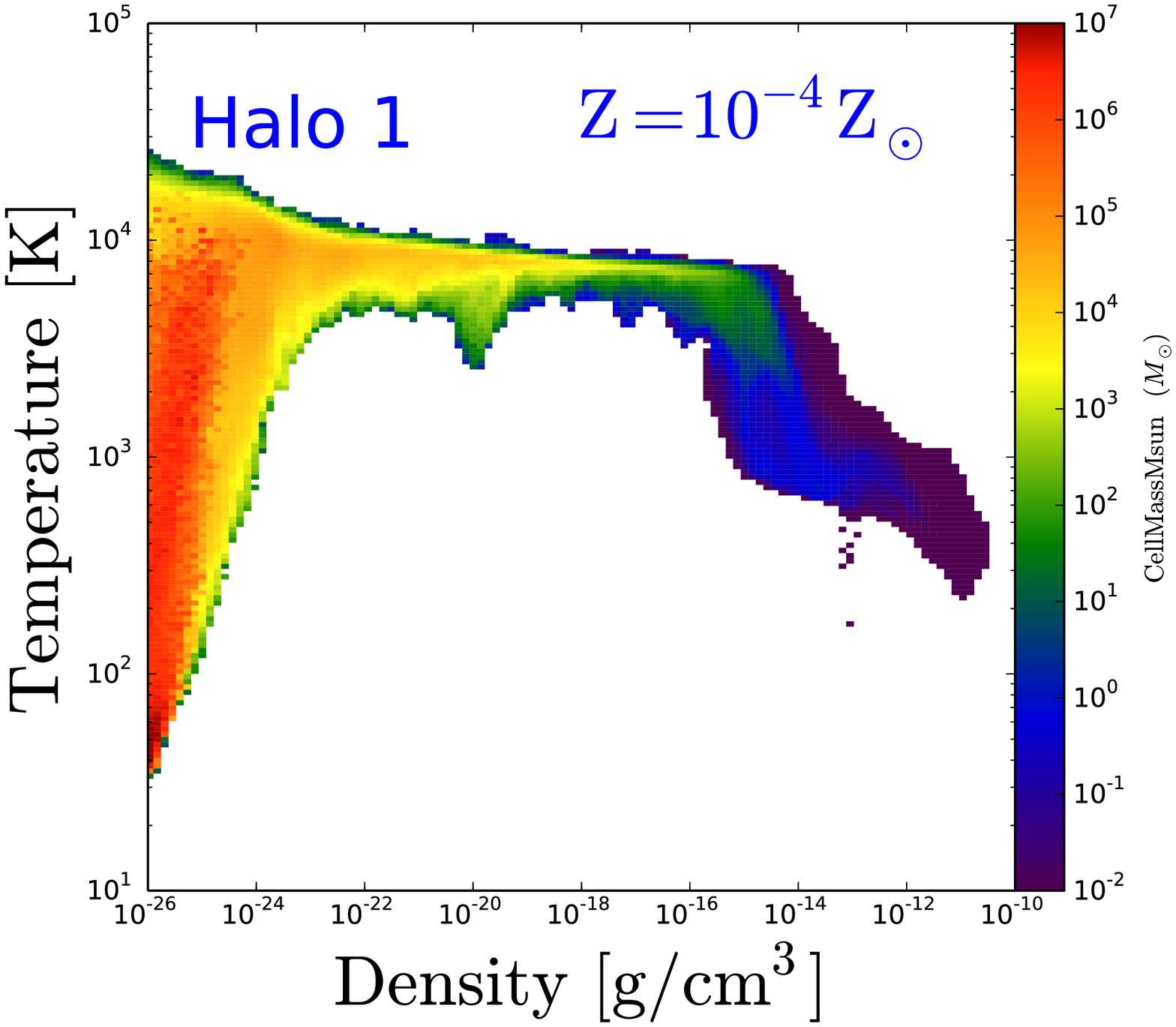}  
\hspace{1.9 cm}\includegraphics[scale=0.39]{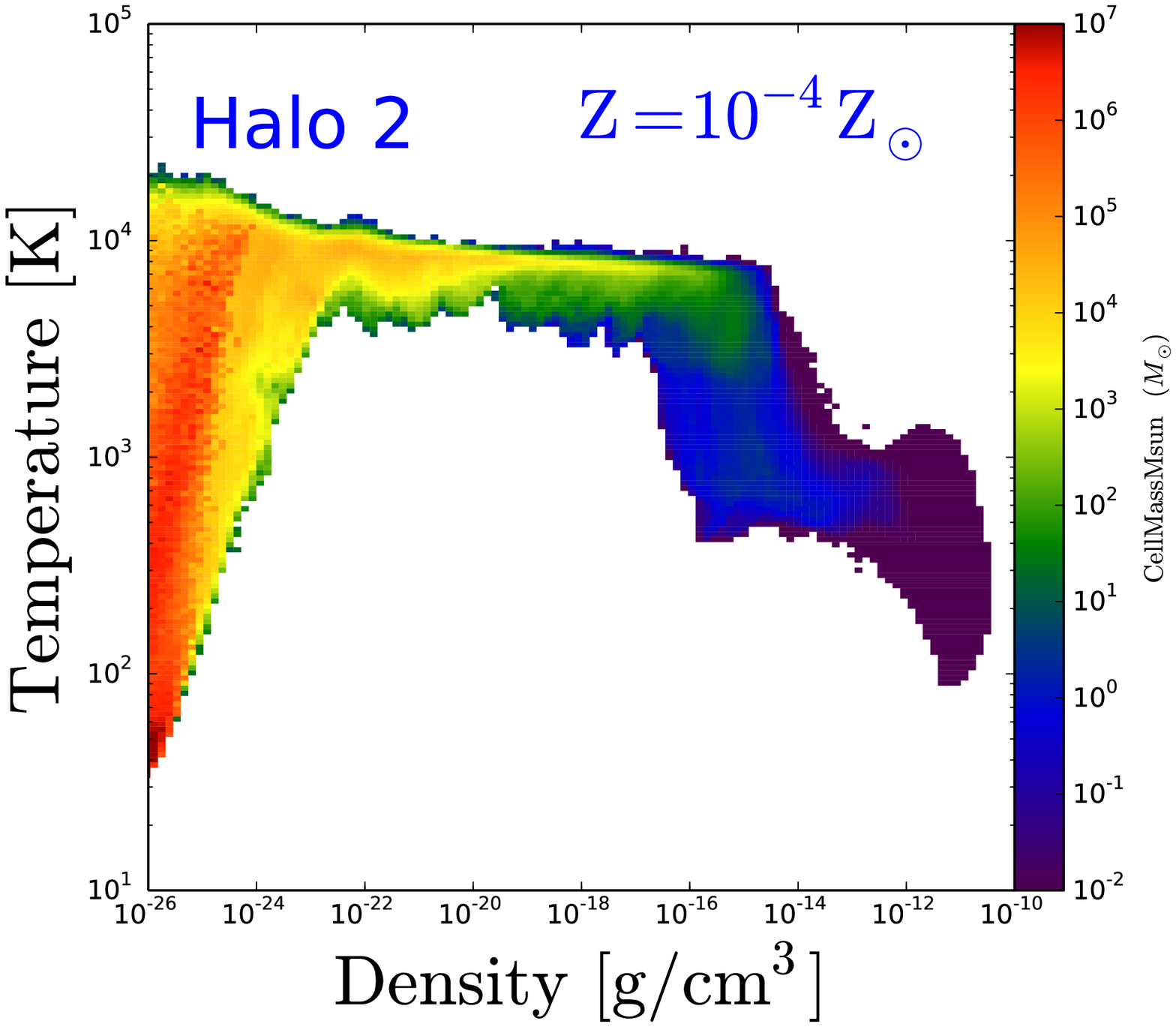} 

\caption{ Density-temperature phase diagram for $\rm Z/Z_{\odot} =10^{-6}, 10^{-5} ~and~ 10^{-4}$  from top to bottom, respectively. The  left panel represents halo 1 and the right panel halo 2. The gas is initially heated up to a few times $\rm 10^4$ K, collapses isothermally up to the densities of $\rm 10^{-16} ~g/cm^3$ in the presence of a strong LW flux and the dust cooling becomes effective at $\rm \geq 10^{-16} ~g/cm^3$ densities depending on the amount of metallicity. Colour-bars show the amount of gas in solar masses for a given density.}
\label{fig1}
\end{figure*}

The density-temperature phase diagrams  of both halos for $\rm Z/Z_{\odot}=10^{-4}, ~10^{-5} ~and ~ 10^{-6}$ are shown in figure \ref{fig1}. In the presence of a strong LW flux which dissociates $\rm H_2$ molecules, the halos are unable to collapse via molecular hydrogen cooling and  continue to grow via merging and accretion until they reach the atomic cooling limit.  Both halos have recently gone through a major merger and the merger history of these halos is described  in \cite{LatifVolonteri15}. At densities of about $\rm \geq 10^{-24} ~g/cm^{3}$, atomic line cooling becomes effective and cools the gas  to about 8000 K. The collapse proceeds isothermally mainly via atomic hydrogen cooling up to densities  of $\rm 10^{-16}~g/cm^{3}$ and the role of metal line cooling remains negligible. At densities $\rm > 10^{-16}~g/cm^{3}$, dust cooling  comes into play and brings the gas temperature down to a few hundred K depending on the amount of dust.  For $\rm Z/Z_{\odot}=10^{-6}$, the collapse remains isothermal up to densities of about $\rm 10^{-12}~g/cm^{3}$,  at higher densities the dust cools the gas down to a temperature of $\rm \sim$1000 K for halo 2 while both warm and cold phases coexist for halo 1 at densities $\rm \geq 10^{-12}~g/cm^{3}$ due to the local variation in the gas density and collapse velocity.  In the intermediate case of  $\rm Z/Z_{\odot}=10^{-5}$, the dust cooling becomes important  earlier at densities of $\rm  \sim 10^{-14}~g/cm^{3}$ for both halos and decreases the gas temperature down to  about  700 K. 

\begin{figure*}
\hspace{0.6 cm}
\begin{minipage}{5cm}
\vspace{-0.9 cm}
\includegraphics[scale=0.7]{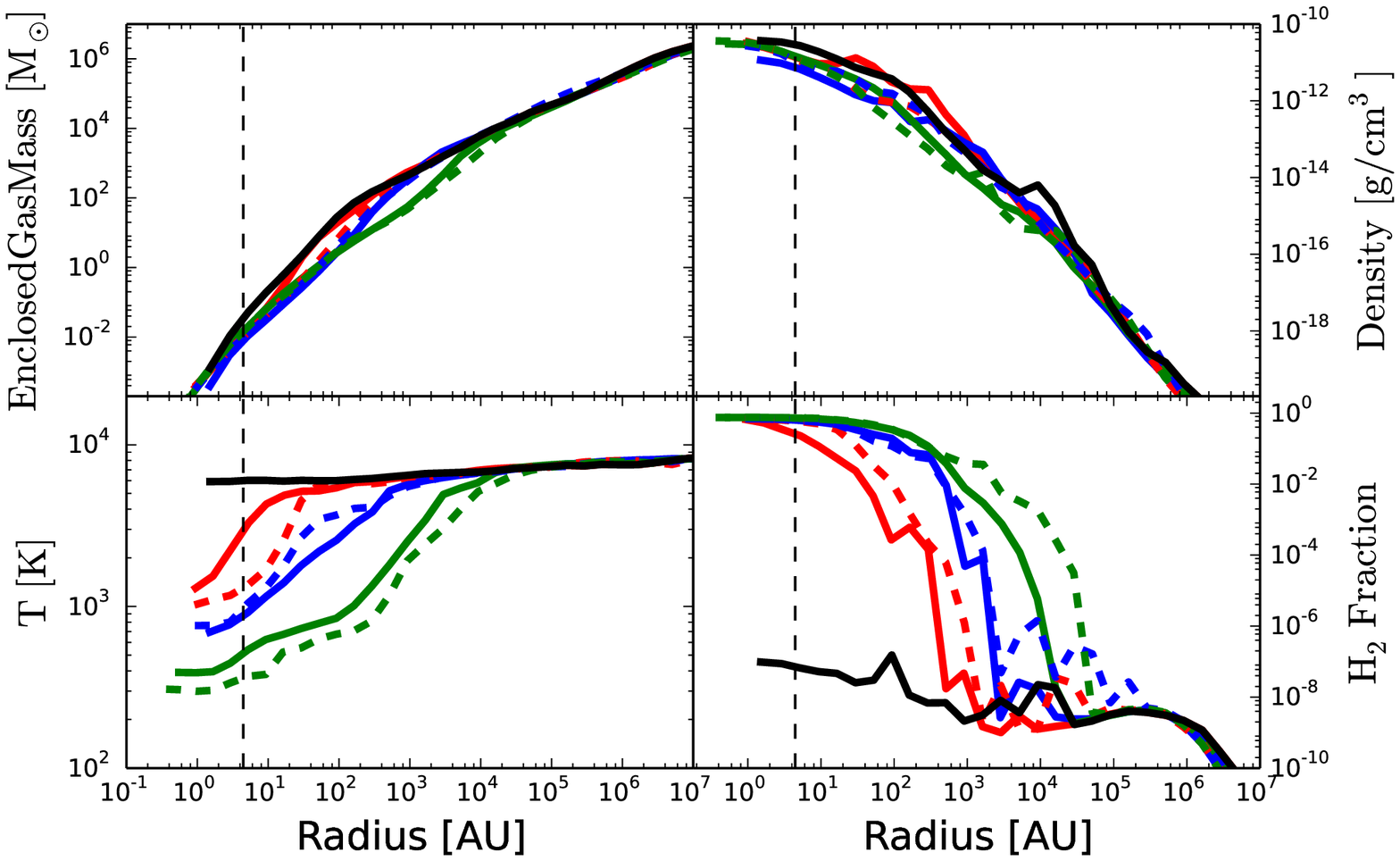} 
\vspace{-1.0cm}
\includegraphics[scale=0.7]{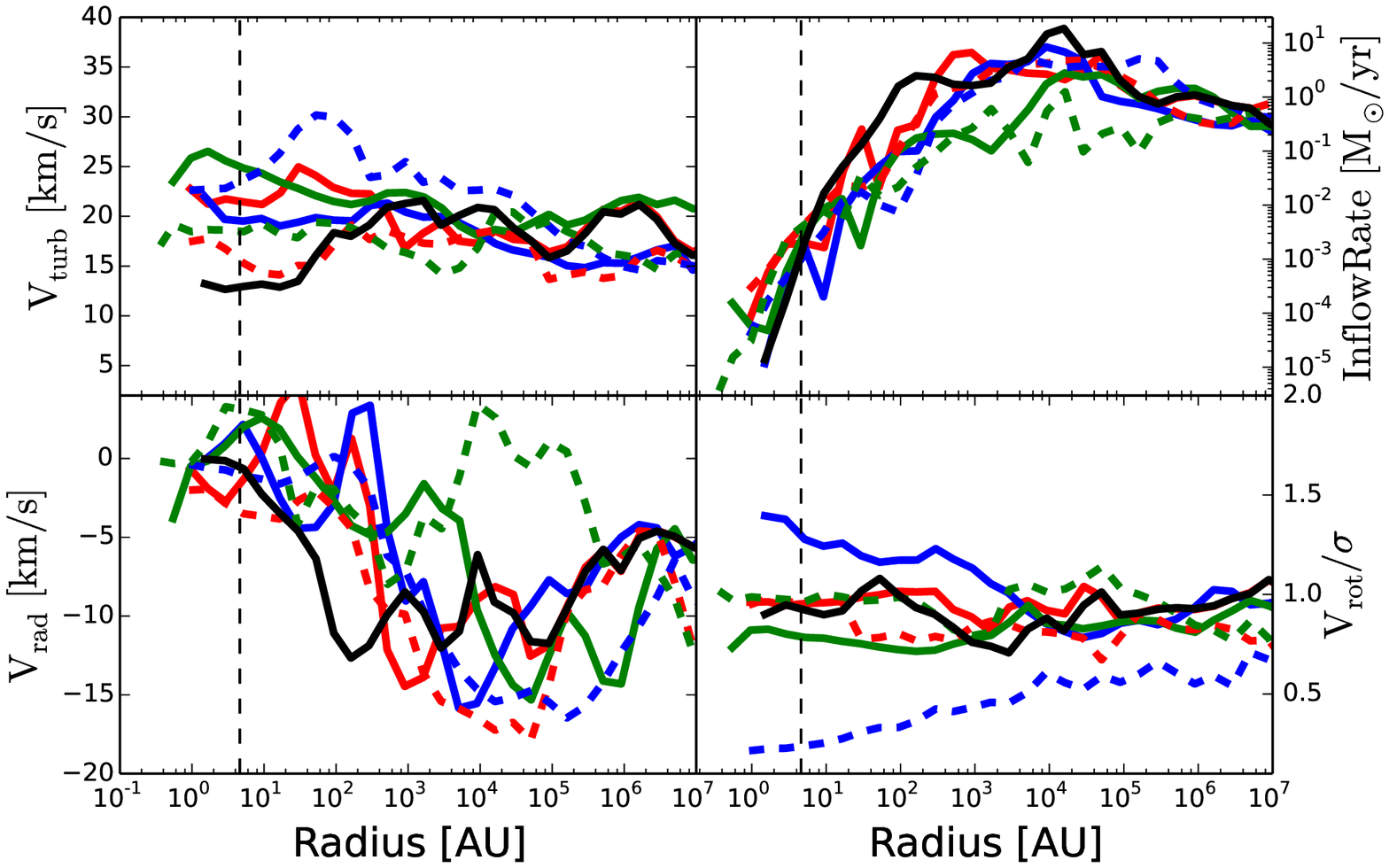} 
\end{minipage}
\caption{Radially averaged and spherically binned profiles of various quantities  for  halo 1 (H1) and  halo 2 (H2) are shown here. The green, blue and red lines represent $\rm Z/Z_{\odot} =10^{-4}, 10^{-5} ~and~10^{-6}$, respectively. The solid lines represent halo 1 and the dashed lines show halo 2. The vertical black dashed line corresponds to the Jeans length.  An isothermal case for halo 1 with zero metallicity is also plotted as a reference and is represented by the black solid line.}
\label{fig2}
\end{figure*}

For $\rm Z/Z_{\odot}=10^{-4}$, the dust grain cooling becomes strong at  densities of $\rm \sim 10^{-16}~g/cm^{3}$ and sharply brings the gas temperature down to a few hundred K. The temperature stalls at about 1000 K at densities between $\rm \sim 10^{-14}-10^{-12}~g/cm^{3}$ due to the chemical heating from the $\rm H_2$ formation on grain surfaces and again the dust cooling takes over at higher densities. These results are in agreement with the one-zone test presented in the previous section and also with  \cite{Omukai2008}. Small differences in the thermal evolution are observed for both halos, the cooling at high densities is generally more effective for halo 2 in comparison with halo 1.  We  show the spherically averaged profiles of density, $\rm H_2$ fraction and temperature in figure \ref{fig2}.  The collapse remains isothermal down to the scales of 10,000 AU irrespective of the metallicity explored in this work due to the strong LW flux. Below this scale dust cooling becomes important and remains confined to the central 10 AU for $\rm Z/Z_{\odot}=10^{-6}$, the inner $\rm \sim$100 AU for  $\rm Z/Z_{\odot}=10^{-5}$ and extends up to 3000 AU for $\rm Z/Z_{\odot}=10^{-4}$. The central temperature is about 300, 700 and 1000 K for $\rm Z/Z_{\odot}=10^{-4}-10^{-6}$, respectively. Similarly, the $\rm H_2$ fraction remains low at scales above $\rm 10^5 AU$ and exceeds  $10^{-3}$ (above which $\rm H_2$ cooling  becomes  important) within the central part of the halo due to its formation on dust grains. For $\rm Z/Z_{\odot}=10^{-4}$,  the $\rm H_2$ abundance gets boosted within the central few 1000 AU while for $\rm Z/Z_{\odot}=10^{-5}$ only in the central few hundred AU and for $\rm Z/Z_{\odot}=10^{-6}$ it remains confined to the central 10 AU.
 
\begin{figure*}
\includegraphics[scale=0.7]{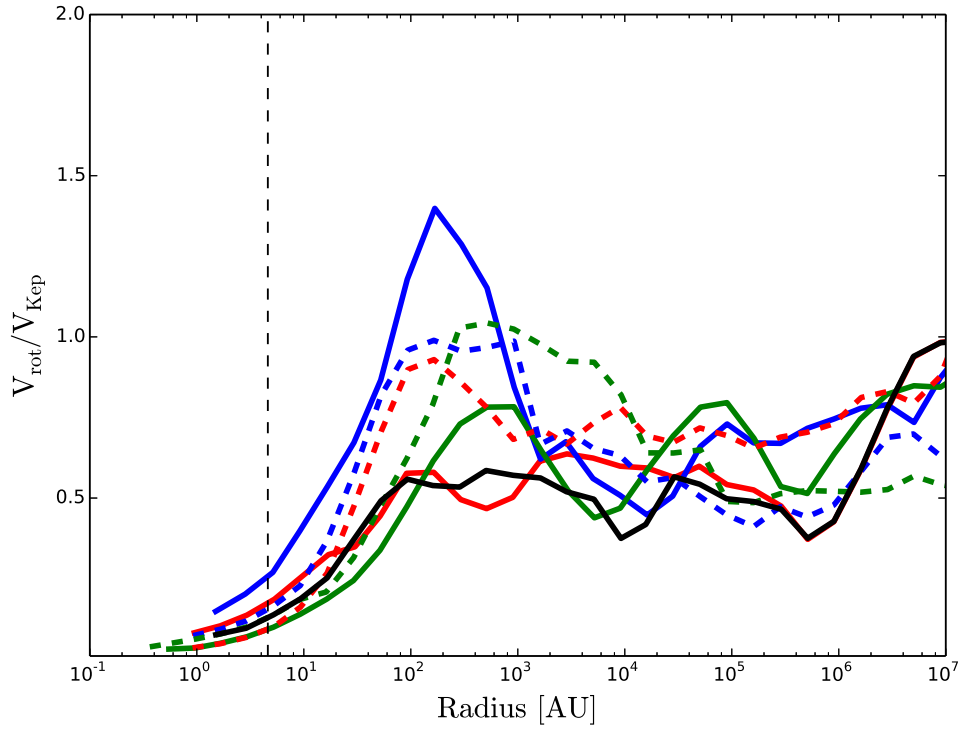} 
\caption{The ratio of rotational to Keplerian velocity  for  halo 1 (H1) and  halo 2 (H2) is shown here. The green, blue and red lines represent $Z/Z_{\odot} =10^{-4}, 10^{-5}$ ~and~$10^{-6}$, respectively. The solid lines represent halo 1 and the dashed lines show halo 2.  An isothermal case for halo 1 with zero metallicity is also plotted as a reference and is represented by the black solid line. The vertical black dashed line corresponds to the Jeans length.}
\label{fig21}
\end{figure*}

 The maximum density reached in our simulations is  about $\rm 3 \times 10^{-11}~g/cm^{3}$ and the density increases with $\rm R^{-1.8}$ above $\rm \sim 100 ~AU$, close to the expected profile from an isothermal collapse (i.e. $\rm R^{-2.0}$). The density profile is shallower for  $\rm Z/Z_{\odot}=10^{-4}$ in the central 1000 AU due to efficient dust cooling while for the other cases it almost follows the isothermal profile. Small bumps in the profile indicate the formation of additional clumps.  The turbulent velocity is about 20 $\rm km/s$ close to the viral radius and increases towards smaller scales up to about 30 $\rm km/s$. This increase in the turbulent velocity is due to the infall of gas towards the center and is higher for dust cooling cases compared to the metal free case. To quantify whether a disk-like structure may form, we computed the ratio of the velocity dispersion (i.e.,  $\rm \sqrt{c_s^2 + v_{turb}^2}$ where $\rm c_s$ is the sound speed and $\rm v_{turb}$ is the turbulent velocity) to the rotational velocity, corresponding to the vertical turbulent support within the disk. The latter provides an estimate for the ratio of vertical scale height to the disk radius. We found that this ratio remains about 1 for all cases except for halo 2 with $\rm Z/Z_{\odot} =10^{-5}$ where V$_{rot}/ \sigma \sim$ 0.3.  This suggests that the rotational velocity is almost comparable to the velocity dispersion and a very thick disk-like structure with H/R of about 1 is formed in all cases. To further estimate the rotational support against gravity, we calculated the ratio of rotational velocity to the Keplerian velocity ($\rm V_{rot}/V_{Kep}$). The ratio of $\rm V_{rot}/V_{Kep}$ is shown in figure \ref{fig21} and is about 0.6 for all cases.  It gets enhanced between 60-1000 AU for dust cooling cases. A peak in $\rm V_{rot}/V_{Kep}$ for  $\rm Z/Z_{\odot}=10^{-5}$ case comes from the density structure inside the halo. Overall, the estimates of $\rm V_{rot}/V_{Kep}$ suggest that the rotational support gets enhanced earlier for higher metallicity cases.  However, enhanced rotational support can delay  the runaway collapse but does not stop it. Moreover, the rotational support is only important in the envelope of the halo and  decreases down within the central core.
 
 
\subsubsection{Mass inflow rates}
Our estimates for  the mass inflow rates ($ 4 \pi R^2 \rho v_{rad}$) are shown in figure \ref{fig2}. Mass inflow rates of $\rm 0.1-1~M_{\odot}/yr$ are observed for $\rm Z/Z_{\odot} \leq10^{-5}$  down to about 100 AU while for $\rm Z/Z_{\odot} = 10^{-4}$  the mass inflow rate starts to decline around $\rm10^5$ AU for  halo 2.  For comparison, we also show a zero-metallicity case where cooling is mainly due to atomic line cooling. We found that  apart from small differences, the inflow rates for $\rm Z/Z_{\odot} \leq10^{-5}$ almost follow the zero-metallicity case. Below 100 AU the mass inflow rates decline in all cases. It comes from the fact that  during the runaway collapse a core-envelope structure develops,  where the core has flat density and  the envelope has a $\rm \sim R^{-2}$ profile, with a core-envelope boundary around 10-100 AU. The radial velocity is high inside the envelope, decreases down in the core and consequently a decline in the instantaneous inflow rate occurs towards the centre. In the central core thermal pressure balances the gravity as the core is not yet gravitationally unstable which results in a decline of the inflow rate.  However,  this may change at later stages when the central core further collapses.  Fragmentation observed in the $\rm Z/Z_{\odot} = 10^{-4}$ cases may also partly contribute  to the decline in the inflow rates. We also note that  mass inflow rate scales with sound speed cube ($\rm \dot{M} \propto c_s^3$) and therefore decreases with the gas temperature. The dust cooling reduces the gas temperature which results in lower inflow rates. The accreting matter inside the envelope will collapse into the protostar at later times and high inflow rates are expected to be maintained. The increase in  the mass inflow rates was observed in our previous simulations  for a metal free case \citep{Latif2013c} as more mass collapsed into the core. We expect a similar trend for other cases until the feedback from the central star blows away the gas or rotational support inhibits the mass accretion.

A similar trend is observed in the radial infall velocity which peaks  around  15 $\rm km/s$ and brings large inflows into the centre of the halo. The radial velocity temporarily becomes positive in a few places due to the presence of clumps, and the positive sign then describes the flow towards the other clump. This is particularly prominent in halo 2 for $\rm Z/Z_{\odot}= 10^{-4}$  and causes a decline in the mass inflow rate. Repeated peaks in the radial velocity indicate the onset of gravitational instabilities which help in  the transfer of angular momentum. The enclosed gas mass in the central 30 pc of the halo is a few times $\rm 10^6~M_{\odot}$.  The mass profile  increases with  $\rm \sim R^{2}$ in the central 100 AU and at larger scales linearly increases with radius. Some differences in the mass profile are observed  between metal free and metal poor cases.  The enclosed mass is a factor of a few lower for  $\rm Z/Z_{\odot}= 10^{-4}$ in comparison with the metal-free case and is probably a consequence of  differences in the thermal structure as well as enhanced rotation for $\rm Z/Z_{\odot}= 10^{-4}$ cases. 

Overall, the mass inflow rates at the end of our simulations seem to be sufficient for forming a supermassive star at-least for $\rm Z/Z_{\odot} \leq 10^{-5}$ while for higher metallicities enhanced rotation as well as fragmentation may limit accretion onto the central object. How long such accretion rates can be maintained depends on whether efficient fragmentation takes place or not.  In the case of efficient fragmentation radiation feedback from in situ star formation  may  shut down accretion onto the central star and a supermassive star cannot form. Even in the absence of fragmentation, an enhanced rotation can limit mass accretion as found for $\rm Z/Z_{\odot} \leq 10^{-4}$ in both halos. We  discuss these possibilities in the section below.

\subsubsection{Fragmentation} 
\begin{figure*}
\hspace{0.7 cm}
\includegraphics[scale=0.3]{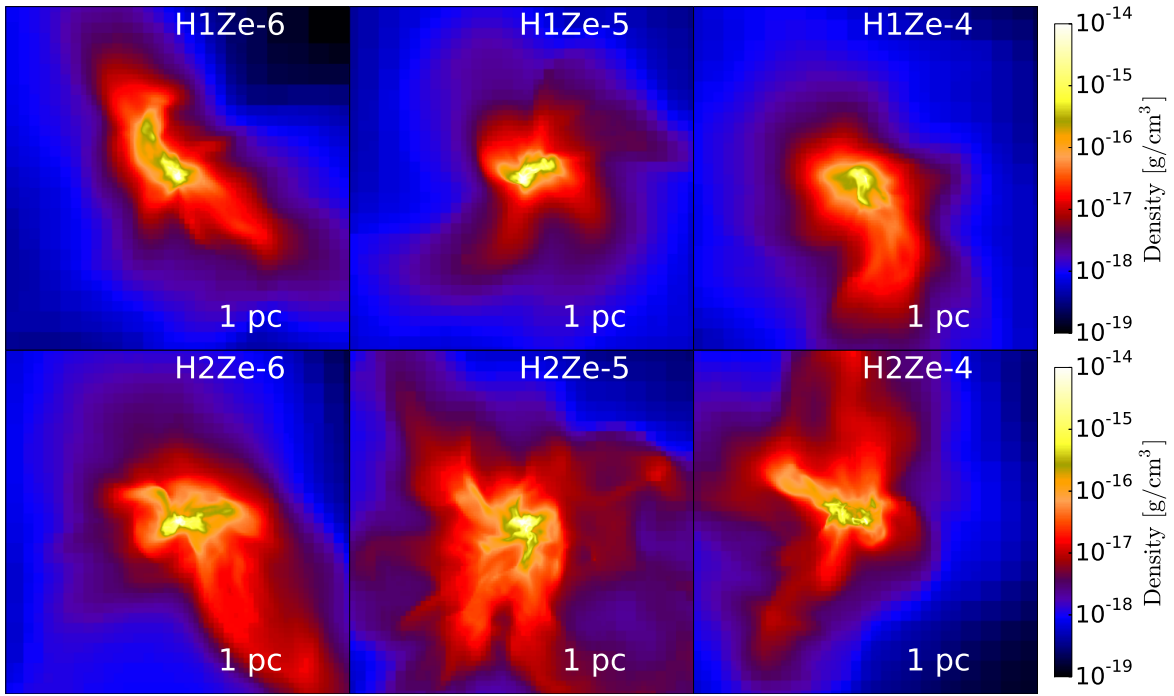} 
\caption{Average gas density along x-axis for the central 1 pc of a halo.  Each  row represents a halo (halo 1(H1) on top and halo 2 (H2) on bottom) and each column represent metallicity (increasing from left to right). H1Ze-6, H1Ze-5 and H1Ze-4 represent $\rm Z/Z_{\odot}=10^{-6},~10^{-5}~and ~10^{-4}$ for halo 1, respectively. Similarly,  H2Ze-6, H2Ze-5 and H2Ze-4 represent $\rm Z/Z_{\odot}=10^{-6},~10^{-5}~and~10^{-4}$ for halo 2.}
\label{fig3}
\end{figure*}

\begin{figure*}
\hspace{0.7 cm}
\includegraphics[scale=0.3]{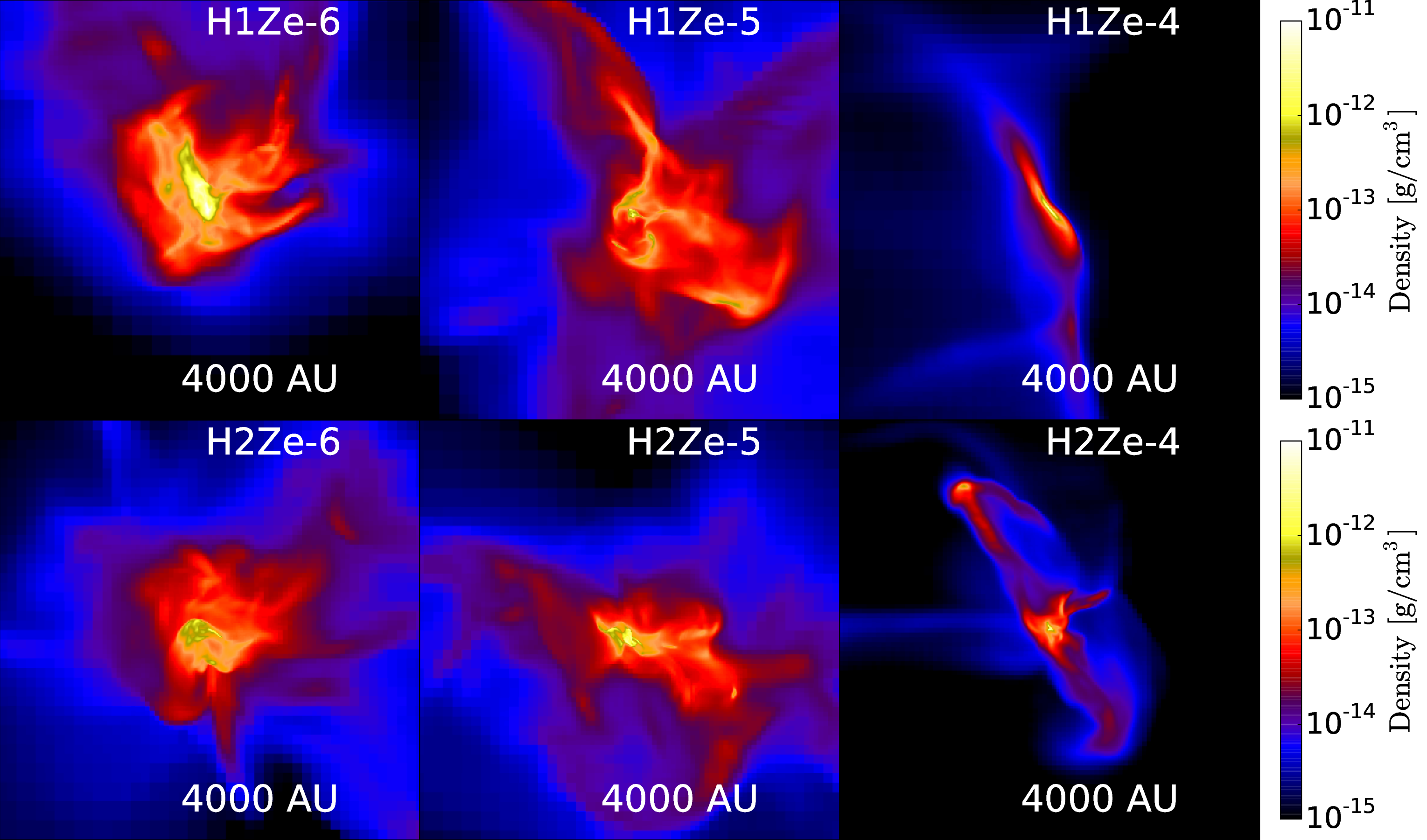} 
\caption{Average gas density along x-axis for the central 4000 AU of a halo.  Each  row represents a halo (halo 1 on top and halo 2 on bottom) and each column represent metallicity (increasing from left to right).}
\label{fig6}
\end{figure*}

To quantify the fragmentation, we  use the YT clump finder \citep{Turk2011} to identify clumps. Our clump finding algorithm locates topologically disconnected structures. The clumps are considered to be bound if the sum of kinetic and thermal energy is less than the potential energy, see \cite{Smith2009}. The density structure within the central 1 pc of the halo is shown in the figure \ref{fig3} for all cases. There is not much substructure found  at this scale as the collapse  is almost isothermal and dust cooling starts to become important around a few 1000 AU. Zooming into the central 4000 AU (see figure \ref{fig6}), significant differences in the morphology of the halos are observed. Particularly, for $\rm Z/Z_{\odot}= 10^{-4}$, the structure is more elongated and collapses into  a filament due to the more efficient cooling compared with other cases where central gas clouds are more spherical. For $\rm Z/Z_{\odot}= 10^{-4}$, two well separated gravitationally bound clumps of about a solar mass are formed in halo 2 while in halo 1 the central clump is of similar mass, gravitationally bound but the second clump is  gravitationally unbound and has a sub-solar mass.  In contrast to this, the clouds are more spherical (although not completely)  for $\rm Z/Z_{\odot}= 10^{-6}$,  very similar to a zero metal case. 

\begin{figure*}
\hspace{0.4 cm}
\includegraphics[scale=0.3]{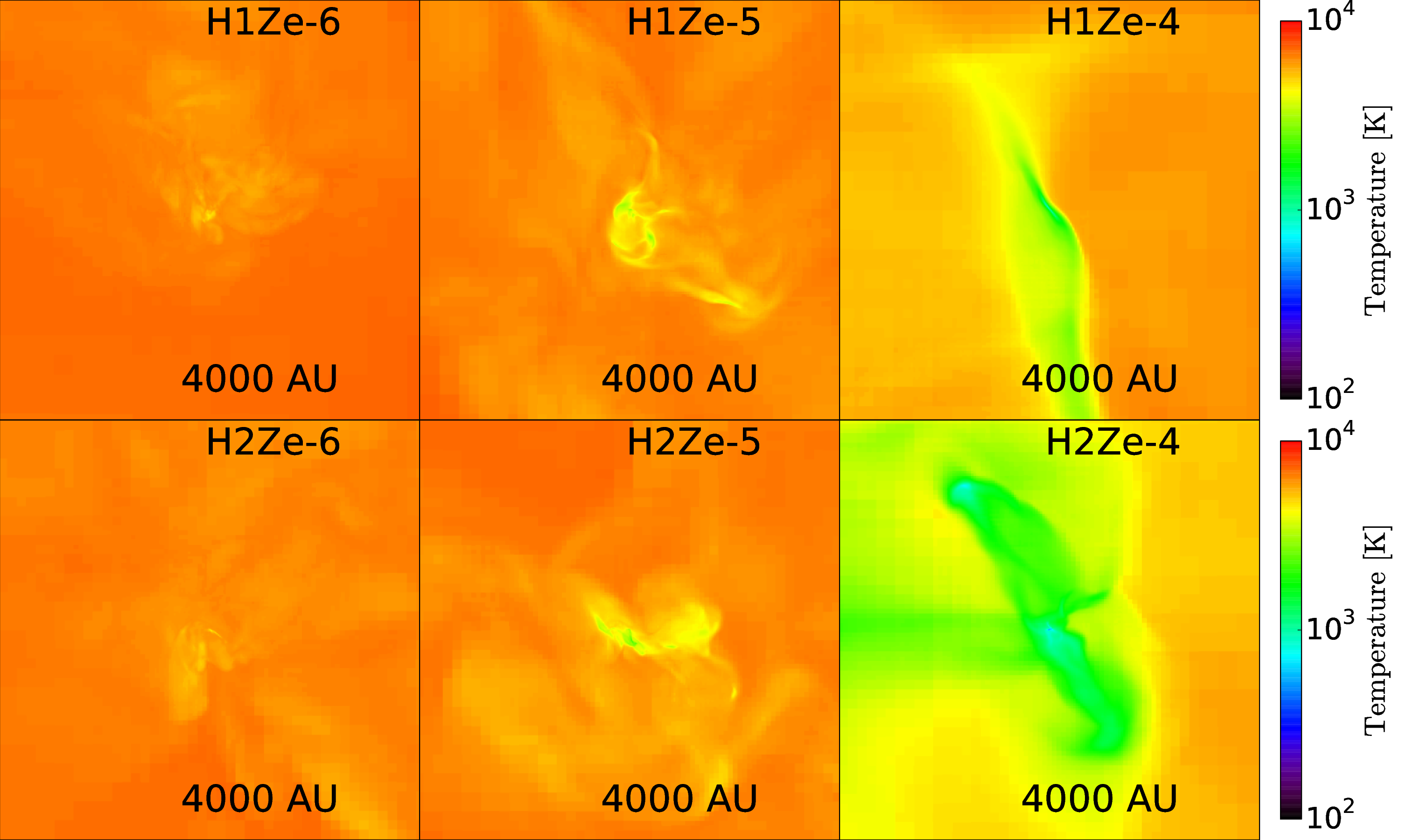} 
\caption{Average temperature weighted by the gas density  along x-axis for the central 4000 AU of a halo.  Each  row represents a halo  (halo 1 on top and halo 2 on bottom) and each column represent metallicity (increasing from left to right).}
\label{fig9}
\end{figure*}

\begin{figure*}
\hspace{-6.0cm}
\centering
\begin{tabular}{c c}
\begin{minipage}{6cm}
\hspace{0.7 cm}
\includegraphics[scale=0.45]{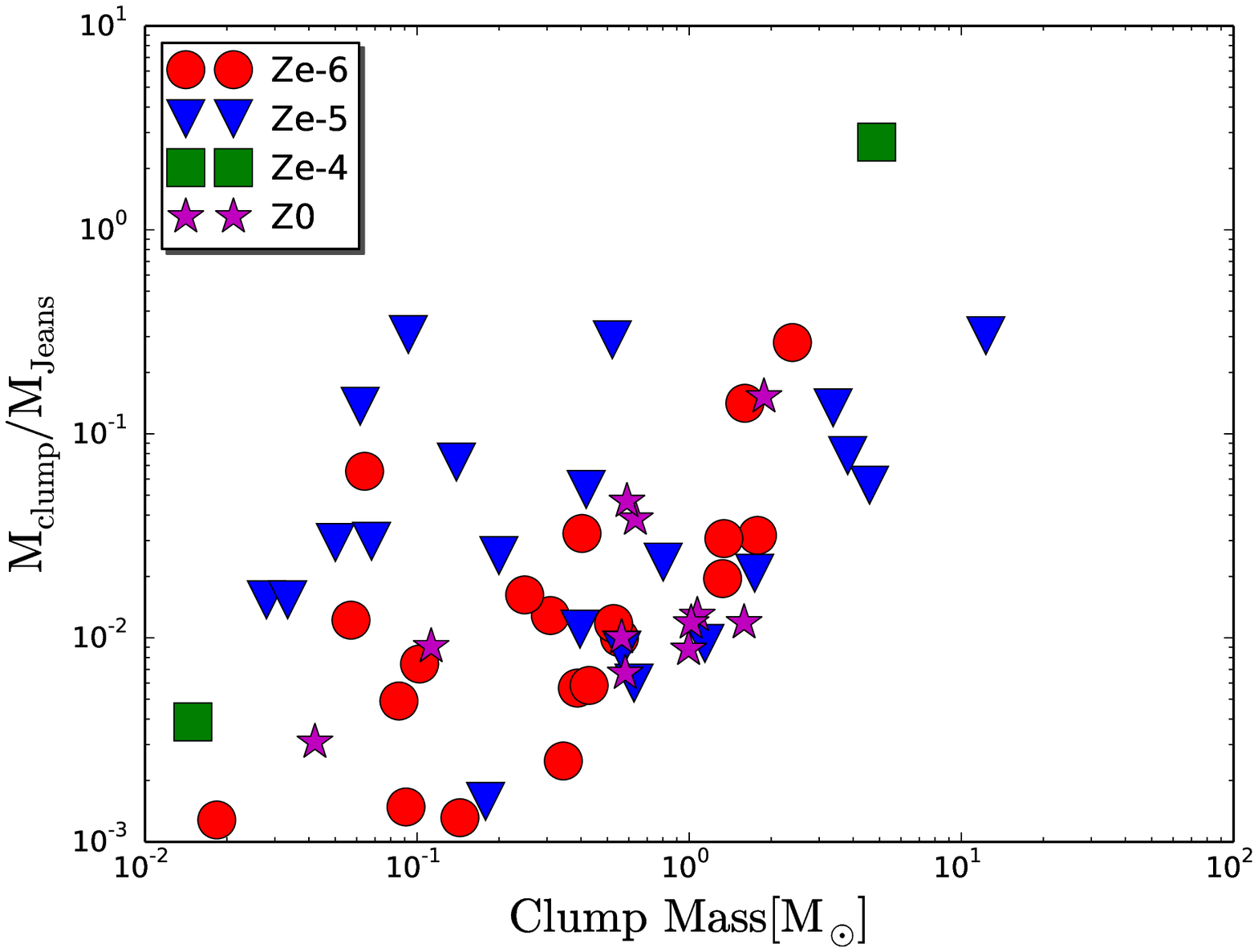}
\end{minipage} &
\begin{minipage}{6cm}
\hspace{2.9 cm}
\includegraphics[scale=0.45]{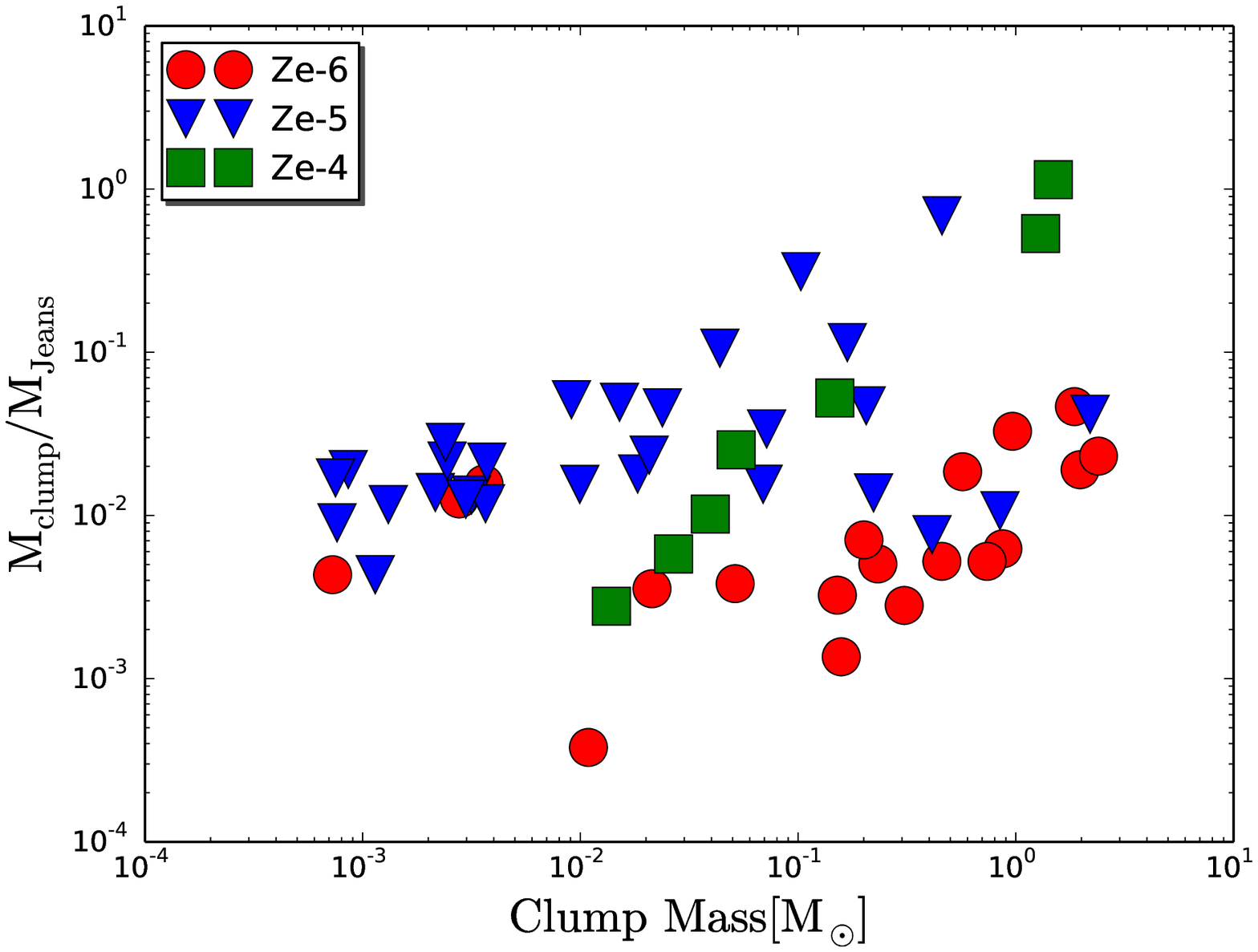}
\end{minipage}
\end{tabular}
\caption{The ratio of clump mass to the Jeans mass is plotted against the clump masses for various metallicities. The left panel halo 1 and the right panel halo 2.  The red symbols represent the clumps forming for $\rm Z/Z_{\odot} = 10^{-6}$, the blue for $\rm Z/Z_{\odot} = 10^{-5}$ and the green for $\rm Z/Z_{\odot} = 10^{-4}$. The clumps  for $\rm Z/Z_{\odot} \leq 10^{-5}$ are gravitationally unbound while the central clumps for $\rm Z/Z_{\odot} = 10^{-4}$ cases are gravitationally bound and more massive.  In general, the ratio of clump mass to the Jeans mass increases with clump masses.}
\label{fig12}
\end{figure*}

For $\rm Z/Z_{\odot}= 10^{-5}$, there is more substructure in the central part of halo 1 with the possibility of multiple clump formation while halo 2 looks similar to the metal free case. Overall the clouds are more elongated compared to the  $\rm Z/Z_{\odot}= 10^{-6}$. We also ran a clump finder for these cases and found that all clumps in these cases are gravitationally unbound and mostly with sub-solar masses as shown in figure \ref{fig12}.  The figure \ref{fig9} shows that only small pockets of gas are cooled by dust cooling. The $\rm Z/Z_{\odot}= 10^{-6}$ case looks very similar to the  metal free case,  the temperature of the dense clumps of gas is about a thousand K  for $\rm Z/Z_{\odot}= 10^{-5}$ and it declines down to a few hundred K for $\rm Z/Z_{\odot}= 10^{-4}$.  Similar differences in the density structure of a gas cloud are observed  by \cite{Peters2012} employing an equation of state with  polytropic index  ($\gamma <$)1. 

We estimated the maximum scale of thermal instability (i.e. $l_{\rm th}= {\rm c_s t_{cool}}$) following the criterion given in \cite{Inoue2015} and found that it is comparable to the Jeans length. It shows that the resolution in our simulations is enough to resolve the thermal instability. Dust cooling triggers the Jeans instability which results in the formation of multiple clumps.  In a  self-gravitating cloud, turbulence can also induce fragmentation by locally compressing the gas as observed in an isothermal collapse, see \cite{Latif2013c}. Overall,  central clumps forming  for $\rm Z/Z_{\odot}= 10^{-4}$ are gravitationally bound and for low metallicities all the clumps are gravitationally unbound. We caution the reader that gravitationally unbound clumps  at the initial stages of the collapse do not reflect  the true amount of fragmentation. To further estimate the fate of these clumps, we followed the collapse of halo 2  with  $\rm Z/Z_{\odot}= 10^{-5}$  to densities of $\rm 10^{-8}~g/cm^{3}$ by employing 5 additional refinement levels. In this case, clumps eventually collapse into the centre of the halo. However, the possibility of fragmentation at later stages of the collapse cannot be ruled out and it may vary from halo to halo. Evolving these simulations further in time becomes  computationally extremely expensive due to the large dynamical range covered in these simulations which results in very short time steps.  In the following section, we discuss the possible implications of efficient fragmentation if it occurs at later times.

\subsection{Fraction of metal enriched halos}
\begin{figure*}
\includegraphics[scale=0.7]{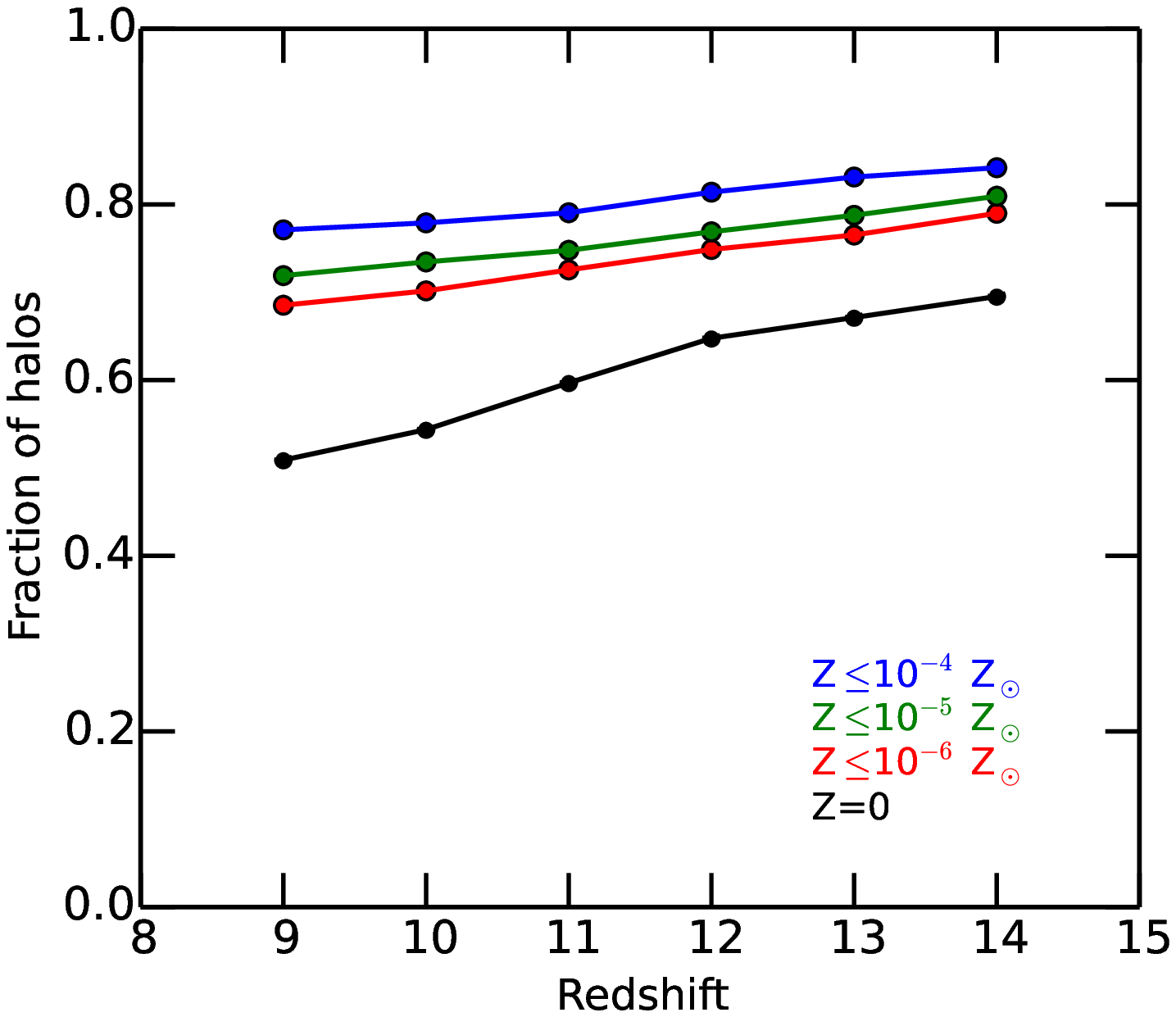} 
\caption{Fraction of halos with metallicities below a given value (see legend) as a function of redshift. In this figure, we  only show the fraction of halos with masses between $\rm 2 \times 10^7 -10^8~M_{\odot}$ in a computational box  of size 10 Mpc for  $\rm Z/Z_{\odot}  \leq 10^{-4}$. The fraction of halos with $\rm Z/Z_{\odot} \leq 10^{-5}$ is about 1.5 times higher than the metal-free halos.}
\label{fig13}
\end{figure*}

We estimate the fraction of metal polluted halos with $\rm Z/Z_{\odot} \le 10^{-4}$ from cosmological hydrodynamical simulations \cite{Habouzit2016}. The computational volume has a  size of 10 Mpc and simulations include recipes for star formation and supernova feedback, a detailed discussion is provided in \cite{Habouzit2016}.  We identified the halos with mass range between $\rm 2 \times 10^7-10^8~M_{\odot}$ using the HaloMaker code \citep{Tweed2009} at least resolved by 100 hundred DM particles and computed their mean metallicity.  Our estimates for the  fraction of halos  with $\rm Z/Z_{\odot} \leq 10^{-4}$ are shown in figure \ref{fig13}. The number of metal polluted halos increases with decreasing redshift  as expected and  the fraction of halos polluted with $\rm Z/Z_{\odot} \leq 10^{-5}$ is about a factor of 1.5 higher in comparison with metal free halos. 

 The halos polluted by a trace amount of metals  can be  irradiated by a strong LW flux if they form in the close vicinity of a star forming galaxy. Our simulations do not self-consistently take into account the local variations in the LW flux required for the direct collapse sites and the simulations volume is not large enough to capture rare sources which provide a strong LW flux of $\rm 10^5~in ~terms ~of ~J_{21}$. Assuming that the metallicity distribution is however uncorrelated with fluctuations in the LW background, if a direct collapse is feasible for metallicities up to $\rm 10^{-5}$, the resulting abundance of the halos is increased by a factor of 1.5 compared to the purely primordial case. 

\section{Implications for black holes formation}
Stellar evolution calculations suggest that the main requirement for the formation of direct collapse black holes are mass inflow rates of $\rm \geq 0.1~M_{\odot}/yr$. For such accretion rates, stellar evolution differs from normal stars, as the stellar radius  increases with mass and no strong UV feedback is produced by the protostar.  Under these conditions, a supermassive star of about  $\rm 10^5~M_{\odot}$ can be formed  which may later collapse into a black hole while retaining most of its mass \citep{Hosokawa12,Hosokawa2013,Schleicher13}. 

In this study, we explored the impact of dust cooling occurring at high densities in the presence of a strong LW flux for two halos of a few times $\rm 10^7~M_{\odot}$  and  metallicities of  $\rm Z/Z_{\odot}= 10^{-4}-10^{-6}$. We resolved the collapse down to sub-AU scales and densities of about a few times $\rm 10^{-11} ~g/cm^{3}$. The presence of large mass inflow rates of  $\rm \geq 0.1~M_{\odot}/yr$ suggests that  trace amounts of metals and dust do not inhibit the formation of massive objects. At least for $\rm Z/Z_{\odot} \leq 10^{-5}$, the mass accretion rates are comparable with an atomic cooling case and no strong fragmentation is observed. It should be noted that the conditions  found here are different from minihalos cooled by dust cooling in the absence of  a strong LW background. Both  mass inflow rates and available gas mass are  about two orders of magnitude higher compared to minihalos. Due to the extremely large spatial range covered in our simulations, the time step becomes very short and does not allow us to evolve simulations  long enough to assess fragmentation and inflow rates at later times. In the following, we consider two possible outcomes  assuming that efficient fragmentation occurs at later times.

In the first case, we consider that  efficient fragmentation  occurs at later times but clumps move inward due to the short migration time scale and merge with the central clump  \citep{Inayoshi2014b,Latif2015Disk,LatifViscous2015,Schleicher2015}.  In the presence of the large inflow rates of $\rm 0.1~M_{\odot}/yr$ found here, viscous heating may become important in the interior of the disk, it will heat the gas up to a few thousand K. Therefore, viscous heating may  suppress fragmentation by increasing the thermal Jeans mass and may help in the formation of massive objects  \citep{LatifViscous2015,Schleicher2015}.  Moreover, for such large inflow rates the impact of UV feedback from  stars is  expected to be quite weak.  Even if some of the clumps survive  or get ejected via 3-body processes and form low mass stars,  a massive central object is still expected to form due to the  clumpy mode of accretion and the stabilisation of the disk due to the viscous heating.  We argue that in such a scenario, massive seed black holes of $\rm  \sim 10^4 ~M_{\odot}$ can form for  $\rm Z/Z_{\odot} \leq 10^{-5}$.

For  the second case, we assume  that efficient fragmentation takes place due to the dust cooling at high densities. Then the clumps are unable to migrate inward and a  stellar cluster forms. A dense stellar cluster is expected to form in such a scenario as dust cooling becomes effective only in the central few 100-1000 AU surrounded by hot gas with a temperature around 8000 K.  The minimum Jeans mass at high densities where dust cooling operates in our simulations is about $\rm > 0.1~M_{\odot}$ and the enclosed mass within a radius of a few hundred AU is about  $\rm 1000~M_{\odot}$. As a consequence of gravitational collapse, a higher gas mass will collapse in the centre over time and may reach up to $\rm 10^6~M_{\odot}$  which is about  few \% of the halo mass as often observed in numerical simulations \citep{Johnson2009,Latif2011}. For a star formation efficiency of about 10 \%, a stellar cluster of about $\rm 10^5~M_{\odot}$ may form. The expected radius of such a stellar cluster is about 1000 AU.  Various studies suggest that  such dense clusters may collapse into a massive black hole either via relativistic instabilities or stellar dynamical processes \citep{Baumgarte1999,PZwart1999,Omukai2008,Devecchi2009,Devecchi2012}. We estimate the mass of a seed black hole forming from the core collapse of a cluster in the following way \citep{Zwart2002},
\begin{equation}
 M_{BH} = m_{*} + 4 \times 10^{-3} f_{c} M_{c0} \gamma ln \Lambda_C ~,
\label{eq2}
\end{equation}
where $\rm m_{*}$ is the mass of a massive star in the cluster which initiates runaway growth, $f_c$ is the effective fraction of binaries formed dynamically,  $\rm M_{c0}$ is  the mass of  the cluster at its birth, $ln \Lambda_c$ is the Coulomb logarithm and $\gamma  \sim 1$, is the ratio of various time scales, see \cite{Zwart2002}. For a star cluster of $\rm M_{c0}= 10^4 ~M_{\odot}$, $\rm m_{*} =100~M_{\odot}$,  $\rm f_c =0.2$, $ ln \Lambda_C =10$, we get  $\rm M_{BH} \sim 180 ~M_{\odot}$. For a $\rm M_{c0} = 10^5 ~M_{\odot}$, expected black holes mass  is $\rm \sim 900 ~M_{\odot}$. Therefore,  depending on the initial mass of a cluster and the central star, massive black holes seeds of up to a thousand solar masses can be formed in metal-poor halos illuminated by the strong LW flux.

\section{Discussion and conclusions}
We have performed high resolution cosmological simulations to  study the impact of trace amounts of  metals and dust cooling in massive primordial halos irradiated by the strong LW flux. To accomplish this goal, we selected two massive halos  of a few times $\rm 10^7~M_{\odot}$  at $z>10$ with metallicities of  $\rm Z/Z_{\odot} = 10^{-4}-10^{-6}$ and  turned on a strong LW radiation of strength $\rm 10^5 ~in ~terms ~of ~J_{21}$. Our simulations cover a large dynamical range by  resolving the collapse starting from cosmological scales down to scales of sub-AU. To take into account the effect of dust and metal line cooling, we extended our previous chemical model and included $\rm H_2$ formation on dust grains, dust grain cooling and heating, CII and OI metal lines cooling following \cite{Omukai2008}.

Our results show that even in the presence of a trace amount of metals and dust, the collapse proceeds isothermally  with temperatures around 8000 K up to densities of about $\rm 10^{-16}~g/cm^{3}$.  Dust cooling becomes effective at densities between  $\rm 10^{-16}-10^{-12}~g/cm^{3}$ and brings the gas temperature down to 100-1000 K for  $\rm Z/Z_{\odot} = 10^{-4}-10^{-6}$, respectively. As expected, dust cooling is more efficient and  occurs at earlier stages of the collapse for higher metallicities in comparison with a lower metallicity case.  In contrast to the isothermal case, $\rm H_2$ formation on dust grains takes place  in the core of the halo and  its fraction gets boosted. For the metal poor cases studied here, metal line cooling is not important but could be effective at metallicities higher than explored here. 

We found that large inflow rates of $\rm 0.1~M_{\odot}/yr$  are available for $\rm Z/Z_{\odot} \leq 10^{-5}$ while for the higher metallicity case  the inflow rates start to decline earlier.  The decline in accretion rate for higher metallicity cases is due to the lower gas temperature in the presence of dust cooling as well as  because of the fragmentation. No gravitationally bound clumps are found for $\rm Z/Z_{\odot} \leq 10^{-5}$ by the end of our simulations while for  $\rm Z/Z_{\odot} = 10^{-4}$ clumps are gravitationally bound and more massive. The  morphology of the central gas cloud is significantly different in $\rm Z/Z_{\odot} = 10^{-4}$, as the cloud is more elongated compared to other cases probably due to the more efficient dust cooling.  \cite{Larson1969} and \cite{Penston1969} found a self-similar solution  under isothermal conditions and later \cite{Yail1983} showed that it changes with the equation of state by including a dependence of the collapse time on the equation of state parameter $\gamma = 1 + \frac{dog \rho}{d log T}$. The latter suggests that the collapse dynamics will be changed by adding dust cooling.  Overall, no strong fragmentation is observed for  $\rm Z/Z_{\odot} \leq 10^{-5}$ cases although the possibility of fragmentation at later stages of the collapse cannot be ruled out. Due to the computational constraints, we were unable to evolve simulations for longer times and therefore cannot make strong statements about fragmentation occurring at later times. 

The presence of large inflow rates suggests that  a massive central star is expected to form  for $\rm Z/Z_{\odot} \leq 10^{-5}$  where dust cooling is confined to the central 10-100 AU depending on the metallicity.  Even if the cloud fragments at later stages of the collapse, the clumps are expected to migrate inward and merge  in the centre as proposed by theoretical works  \citep{Inayoshi2014b,LatifViscous2015}. Viscous heating which becomes particularly important in the presence of large inflow rates and rapid rotation may further help in suppressing the fragmentation by heating the gas and evaporating the dust grains \citep{LatifViscous2015,Schleicher2015}.  For   $\rm Z/Z_{\odot} = 10^{-4}$, efficient fragmentation is expected to occur  which may lead to the formation of a stellar cluster.

The strength of the LW flux required to keep the collapse completely isothermal in metal-free halos is about a few times $\rm 10^4$ \citep{Latif2015a} and in the present work we employed even a higher value of $\rm 10^5~in ~units ~of ~J_{21}$. The expected  number density of direct collapse black holes forming in  primordial halos for  $\rm  J_{crit} \geq 10^4$ is a few orders of magnitude lower than the observed abundance of quasars at z=6 (i.e. 1 per $\rm Gpc^{3}$ \citep{Willot2010}) extrapolating the results of \cite{Habouzit2015}. Even relaxing the constraint of  primordial gas and allowing the direct collapse black holes to  form in halos with $\rm Z/Z_{\odot} \le 10^{-5}$ in the presence of a strong LW flux  only increases their abundance by a factor of two.  Even under these conditions, the formation sites for the direct collapse black holes remain  rare.

For metallicities higher than explored here, cooling due to  the fine structure lines of CII and OI becomes important at low densities about $\rm 10^{-20}~ g/cm^{3}$ even in the presence of a strong LW flux and cools the gas down to a few 10 K. In such cases, the thermal instability induces fragmentation and determines the mass spectrum of  the resulting star cluster \citep{Inoue2015}.

 \section*{Acknowledgments}
This project has received funding from the European Union's Horizon 2020 research and innovation programme under the Marie Sklodowska-Curie grant  agreement N$^o$ 656428.
The research leading to these results has also received funding from the European Research Council under the European Community's Seventh Framework Programme (FP7/2007-2013 Grant Agreement no. 614199, project ``BLACK''). KO acknowledges the Grant-in-aids from the Ministry of Education, Culture, Sports, Science, and Technology (MEXT) of Japan (25287040). This work was granted access to the HPC resources of TGCC under the allocation x2015046955 made by GENCI. The simulation results are analyzed using the visualization toolkit for astrophysical data YT  \citep{Turk2011}.

 \bibliography{smbhs.bib}
 
\newpage

\end{document}